\documentclass[aps,prb,floatfix,showpacs,showkeys, twocolumn,preprintnumbers,amsmath,amssymb]{revtex4-1}
\usepackage[pdftex]{graphicx}
\usepackage{dcolumn}   % Align table columns on decimal point
\usepackage{bm}        % bold math
\usepackage{color}     % coloring
\usepackage{rotating}  % rotating package
\begin{document}

\title{First-principles studies of orbital and spin-orbit properties of GaAs, GaSb, InAs, and InSb zinc-blende and wurtzite semiconductors}
\author{Martin Gmitra and Jaroslav Fabian}
\affiliation{Institute for Theoretical Physics, University of Regensburg, 93040 Regensburg, Germany}
\begin{abstract}
We employ first-principles techniques tailored to properly describe semiconductors (modified Becke-Johnson potential added to the exchange-correlation functional), to obtain the electronic band structures of both the zinc-blende and wurtzite phases of  GaAs, GaSb, InAs, and InSb. We extract the spin-orbit fields for the 
relevant valence and conduction bands at zone center, by fitting the spin-splittings resulting from the lack of space inversion symmetry of
these bulk crystal structures, to known functional forms---third-order polynomials. We also determine the orientations of the 
spin-orbit vector fields (for conduction bands) and the average spins (valence bands) in the momentum space. We describe
the dependence of the spin-orbit parameters on the cation and anion atomic weights. These results should be useful
for spin transport, spin relaxation, and spin optical orientation modeling of semiconductor heterostructures, as well as for realistic studies of semiconductor-based Majorana nanowires, for which accurate values of spin-orbit couplings are needed.
\end{abstract}
\date{\today}
\maketitle

%-------------------------------------------------------------------
\section{Introduction}
%-------------------------------------------------------------------

Semiconductor spintronics \cite{Zutic2004:RMP, Fabian2007:APS} builds on fundamental aspects of the 
electron spin interactions. Spin-orbit coupling is particularly important, as it links charge and spin, allowing
to control one with the other. There are two main traits of the spin-orbit coupling in the electronic band structure. First, spin-orbit coupling
leads to orbital splittings of the bands, while preserving the spin degeneracy. This originates in the fine
structure of the host atomic orbitals. A well known example is the spin-orbit split-off band gap in zinc-blende 
semiconductors. Second, and this is limited to crystals and structures lacking space inversion symmetry, 
spin-orbit coupling leads to spin splitting of the energy bands. This splitting is an emerging physics due to the
crystal field, without a counterpart in atomic-orbitals physics. Effectivelly, the spin-orbit coupling gives rise
to momentum-dependent spin-orbit fields, in analogy to Zeeman fields. Both zinc-blende and wurtzite crystals
lack space inversion symmetry, and so exhibit spin splittings due to spin-orbit fields. Again, the most known
example is the Dresselhaus field \cite{Dresselhaus1955:PR} in zinc-blende semiconductors, which 
describes a cubic (in momentum)  spin splitting away from zone center. 

Spin-orbit coupling in semiconductors leads to spin relaxation\cite{Fabian1999:JVSTB}, optical spin orientation\cite{Meier1984:book}, 
spin Hall effects\cite{Sinova2015:RMP}, persistent spin structures\cite{Schliemann2016}, or the spin galvanic phenomena.\cite{Ganichev2008:IJMPB} Recently, it has been proposed that spin-orbit fields in semiconductor nanowires with induced proximity superconductivity
can support Majorana bound states.\cite{Lutchyn2010:PRL,Oreg2010:PRL}
Experimentalists are searching for Majorana states in both zinc-blende 
InSb\cite{Mourik2012:Science, Deng2012:NL} and wurtzite InAs\cite{Albrecht2016:Nature,Das2012:NP} nanowires. 
To determine the regime for such states to exist, accurate values for the
spin-orbit fields are required for the underlying semiconductor materials.

Determination of spin-orbit coupling, especially in zinc-blend III-V semiconductors,\cite{Cardona1988:PRB,Winkler2003:book,Chantis2006:PRL} has a long history. Unfortunately, 
there are conflicting values reported in the literature. For example, the Dresselhaus coupling in GaAs is determined 
in the range from 9 to 28~eV$\mathrm{\AA^3}$  (see Table  III.7. in Ref. \onlinecite{Fabian2007:APS}). Experimentally, the difficulty is to have reliable models to extract the parameters, while theoretically one needs reliable electronic band structure calculations. This is especially acute for wurtzite phases which are predominantly found (in several polytypes \cite{Caroff2009:NN})
in nanowires of GaAs \cite{McMahon2005:PRL}, InAs \cite{Kriegner2011:NL,Zardo2013:NL}, and 
InSb.\cite{Kriegner2011:NL} Recently, several groups have demonstrated a controlled growth of nanowires with 
specific lattice structure.
 \cite{Caroff2009:NN, Dick2010:NL, Joyce2010:NL, Krogstrup2010:NL, Shtrikman2009:NL1, Shtrikman2009:NL2, Krogstrup2015:NM} This versatile growth of III-V semiconductor nanowires has opened the possibility to study anisotropic photonic properties in both zinc-blende and wurtzite phases.\cite{Wilhelm2012:NS} For example, 
different microscopic contributions to the spin-orbit coupling result in unusual spin dynamics with anisotropic spin relaxation, as measured by time-resolved micro-photoluminescence on single WZ nanowires.\cite{Furtmeier2016}

The band gap of III-V semiconductors is at the centre of the Brillouin zone and its size decreases with the increasing atomic number of the atomic species. There were several investigations of the electronic structure of III-V semiconductors starting from empirical nonlocal pseudopotantials \cite{Chelikowsky1976:PRB} to modern density functional theory.\cite{Hohenberg1964:PR} The key ingredient of the density functional theory is the exchange-correlation functional which should contain the relevant information about many body interactions. Unfortunately, local and semilocal models for the exchange correlation functional fail to reproduce known band gap values.\cite{Murayama1994:PRB} It has been
shown that the strong underestimation of the fundamental gap in GaAs by the local density approximation \cite{Perdew1981:PRB} leads to a spin splitting parameter that is 14 times greater than the value predicted from a GW 
band structure that results in the correct band gap.\cite{Chantis2006:PRL,Chantis2006:PRL-erratum}

The spin-orbit couplings of III-V semiconductors in wurtzite phases have not been systematically addressed. In this paper we use density functional theory calculations with semilocal exchange modified Becke-Johnson potential \cite{Becke2006:JCHP,Tran2009:PRL} to calculate spin-orbit coupling parameters for the conduction and valence bands in both zinc-blende and wurtzite phases. Our results compare favorably with the existing GW predictions, \cite{Chantis2006:PRL} for the zinc-blende crystals, while predict the spin-orbit fields and spin splittings for the
wurtzite phases, for GaAs, GaSb, InAs, and InSb. We systematically investigate the influence of the atomic numbers of the cations and anions on the spin splittings of the valence and conduction bands, and provide realistic parameters for the functional forms of the spin-orbit fields. The DFT methodology can also be applied to semiconductor slabs, 
as was recently shown in Ref. \onlinecite{Soluyanov2016:PRB}, which studied spin-orbit splittings in confined zinc-blende InSb. We believe that our results provide a useful database and benchmark for more empirical approaches, such as $k\cdot p$ methods\cite{Paulo2016:arXiv}, which can be used to model larger structures such as nanowires.

%-------------------------------------------------------------------
\section{Crystal structure}
%-------------------------------------------------------------------

Binary III-V semiconductors form crystals with tetrahedral coordination, with atoms 
arranged either in zinc-blende (ZB) or wurtzite (WZ) lattice structures.
A ZB crystal comprises two interpenetrating face-centered-cubic (fcc) Bravais lattices,
each of a different atomic species, cation and anion; the corresponding space 
group is $F\bar{4}3m$.
%
%Recently grown III-V nanowires commonly display ZB and wurtzite (WZ) polymorphism \cite{Caroff2009:NN}. 
%
A WZ crystal is constructed from two interpenetrating hexagonal-close-packed (hcp) lattices; the space group is $P6_3mc$. 

\begin{figure}
\centering
\includegraphics[width=0.98\columnwidth]{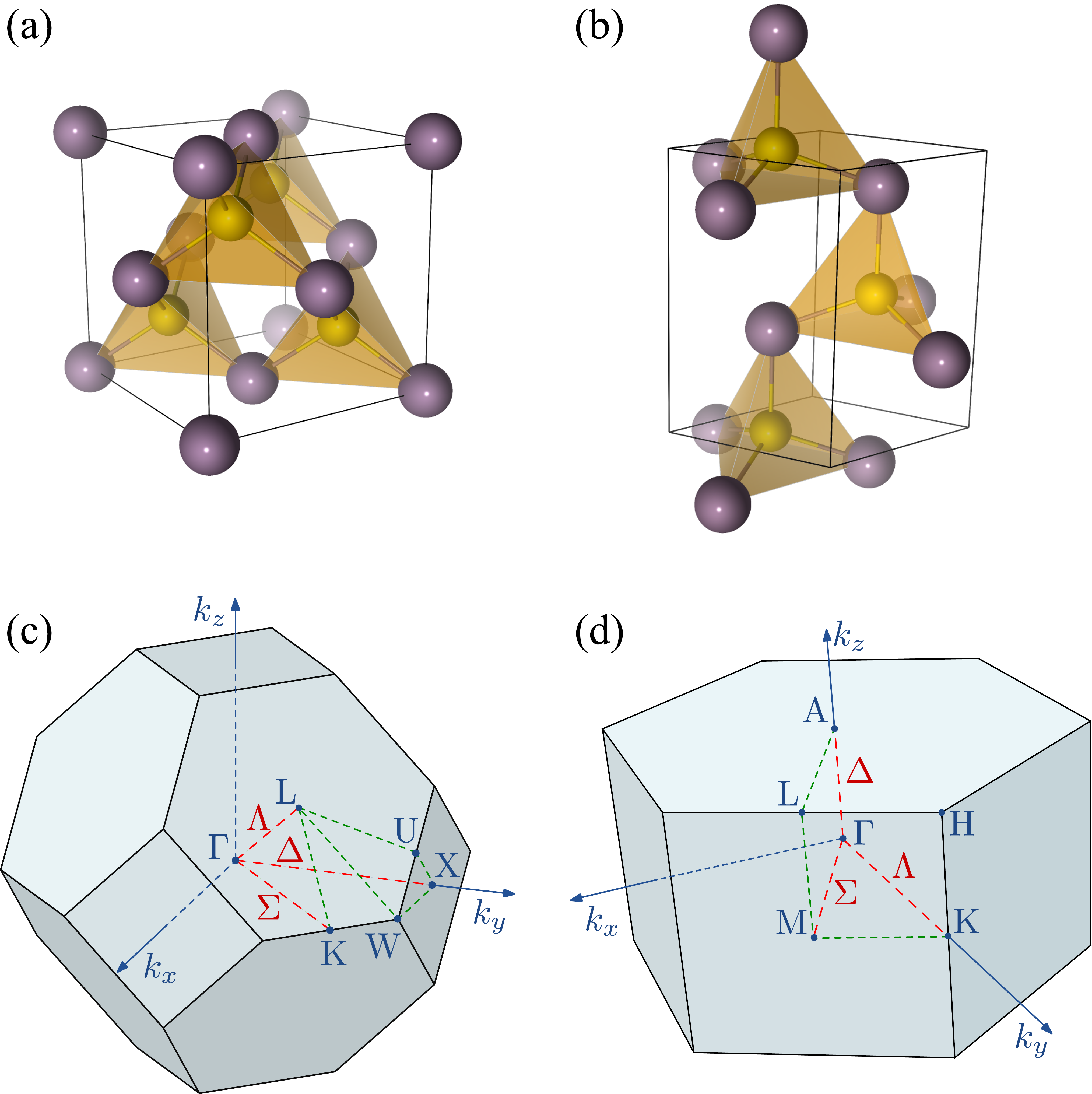}
\caption{\label{Fig:structures}
		Crystal structures of III-V semiconductors:
		(a)~zinc-blende and (b)~wurtzite unit cells.
		First Brillouin zones with labeled high symmetry points and lines for (c)~zinc-blende 
		(truncated octahedron), and (d)~wurtzite (hexagonal prism) crystals.
	}
\end{figure}

The differences between the two structures are manifested by viewing them along the [111] direction for ZB and [0001] for WZ, along which both look like stacked hexagonal layers. The atoms are identical within each layer, and the layers alternate between the anion and the cation. Each anion has four nearest neighbor cations positioned in a tetrahedron. ZB is based on an fcc lattice of anions whereas WZ is derived from an hcp array of anions. The nearest neighbor connections are similar, but the distances and angles to further neighbors differ. In Fig.~\ref{Fig:structures}(a,b) we show the tetrahedra around each anion connecting four cations. For the ZB structure the tetrahedra form triangular lattice planes normal to the [111] direction. The planes are shifted with each other forming an ABC stacking sequence. For the WZ structure the triangular lattice of tetrahedra along [0001] forms an Ab stacking sequence, where the b plane has rotated tetrahedra by 60 degrees with respect to the A plane.
%
%------------------------------------------------------------------
%------------------------------------------------------------------
%
An ideal WZ crystal has the in-plane lattice constant given by $a_\mathrm{WZ} = a_\mathrm{ZB} / \sqrt{2}$. The lattice constant along the $c$ axis (axis perpendicular to the hexagon) is $c = a_\mathrm{WZ} \sqrt{8/3} $. 
%As the WZ crystal is tetrahedral, the nearest neighbors are the same in the two polytypes. In addition, we see from Fig.~\ref{Fig:structures} that nine of the 12 nearest neighbors in WZ are the same as in ZB. 
The first Brillouin zone is a truncated octaheadron for the ZB phase, and a hexagonal prism for the WZ phase, see Fig.~\ref{Fig:structures}(c,d).

The particular order of the cations and anions within the unit cell determines the spin orientation \cite{Cardona1988:PRB} caused by spin-orbit fields. In this work we use the following ordering. Our ZB structure is formed by the Bravais basis vectors $\mathbf{a}_1=a(0,1,1)/2$, $\mathbf{a}_2=a(1,0,1)/2$, $\mathbf{a}_3=a(1,1,0)/2$, with the cation (Ga, In) at (0,0,0) and anion (As, Sb) at (1,1,1)/4; $a$ is the cubic lattice constant. The primitive basis vectors of our hexagonal Bravais lattice of the WZ structure are $\mathbf{a}_1=a(\sqrt{3},-1,0)/2$, $\mathbf{a}_2=a(0,1,0)$, and $\mathbf{a}_3=c(0,0,1)$; $a$ and $c$ are the in-plane and perpendicular lattice parameters. Using the three basis vectors $\mathbf{a}_i$ ($i=1,2,3$) we place
the atoms as follows:
$(2/3,1/3,u)$ and $(1/3,2/3,1/2+u)$ with $u=0$ for anion and $u=3/8$ for cation.
%As: (2/3,1/3,0) and (1/3,2/3,1/2) ???
%Ga: (2/3,1/3,u) and (1/3,2/3,1/2+u) ???
In general, we also allow for the shift $u=3/8+\epsilon$, with a small dimensionless  internal cell structural parameter $\epsilon$ describing a deviation from ideal tetrahedrons as one observes in SiC polytypes \cite{Kackell1994:PRB,McMahon2005:PRL,Belabbes2012:PRB}.

%-------------------------------------------------------------------
\section{Methods}
%-------------------------------------------------------------------

Standard local and semilocal exchange-correlation functionals applied within density functional theory (DFT) typically underestimate the semiconducting gaps. A simple
rigid shift of the bands (scissor operator) would still preserve the wrong dispersion \cite{Cardona1988:PRB} and spin-orbit splittings. The conduction bands have to be 
correctly (as much as possible) located in energy with respect to the valence bands
in order to have realistic descriptions of the spin physics in semiconductors.

Along state-of-the-art GW calculations for III-V semiconductors in zinc-blende  \cite{Chantis2006:PRL,Chantis2010:PRB} and for InAs and GaAs in wurtzite structure \cite{Zanolli2007:PRB,Cheiwchanchamnangij2011:PRB}, there are less computationally demanding studies using local density approximation (LDA)\cite{Panse2011:PRB} and LDA-1/2 method.\cite{Ferreira2008:PRB,Belabbes2012:PRB} In between, on the computational complexity level, are methods of hybrid functionals that mix a portion of the exact exchange with semilocal exchange-correlation functionals. Such methods predict reasonable effective masses and gaps in ZB\cite{Betzinger2010:PRB,Tran2011:PRB,Friedrich2012:JPCM} and WZ structures. \cite{Heiss2011:PRB} However, even for these intermediate techniques (not to mention GW), to resolve the fine energy scales on which spin-orbit coupling is manifested
requires tremendous computational efforts.

An efficient and accurate alternative for the electronic structure calculations of semiconductors by means of DFT provides the modified \cite{Tran2009:PRL} exchange Becke-Johnson (mBJ) potential.\cite{Becke2006:JCHP} It has been shown that the semilocal Becke-Johnson potential makes predictions for the band gaps similar \cite{Kim2010:PRB,Koller2011:PRB,Koller2012:PRB} to hybrid functionals \cite{Kim2009:PRB} and GW methods.\cite{Chantis2006:PRL,Luo2009:PRL,Chantis2010:PRB} 
The semilocal approach is computationally on a par with LDA\cite{Perdew1981:PRB} or PBE\cite{Perdew1996:PRL} calculations. Therefore, it is well suited for
investigating subtle spin-orbit effects, but also for studying extensive systems 
such as surfaces and interfaces including spin-orbit coupling.

%In this work we have calculated the electronic structure by means of density functional theory using modified \cite{Tran2009:PRL} exchange Becke-Johnson (mBJ) potential \cite{Becke2006:JCHP} as implemented in Wien2k code.\cite{wien2k} It has been shown that the semilocal Becke-Johnson potential provides prediction of band gaps of the same order \cite{Kim2010:PRB,Koller2011:PRB,Koller2012:PRB} as hybrid functionals \cite{Kim2009:PRB} or GW method.\cite{Chantis2006:PRL,Luo2009:PRL,Chantis2010:PRB} In addition the semilocal approach is barely expensive as LDA\cite{Perdew1981:PRB} or PBE\cite{Perdew1996:PRL} calculations, therefore, it can be used for study of extensive systems such surfaces or interfaces including spin-orbit coupling.

%
%------------------------------------------------------------------
\begin{figure*}
	\centering
	\includegraphics[width=1.99\columnwidth]{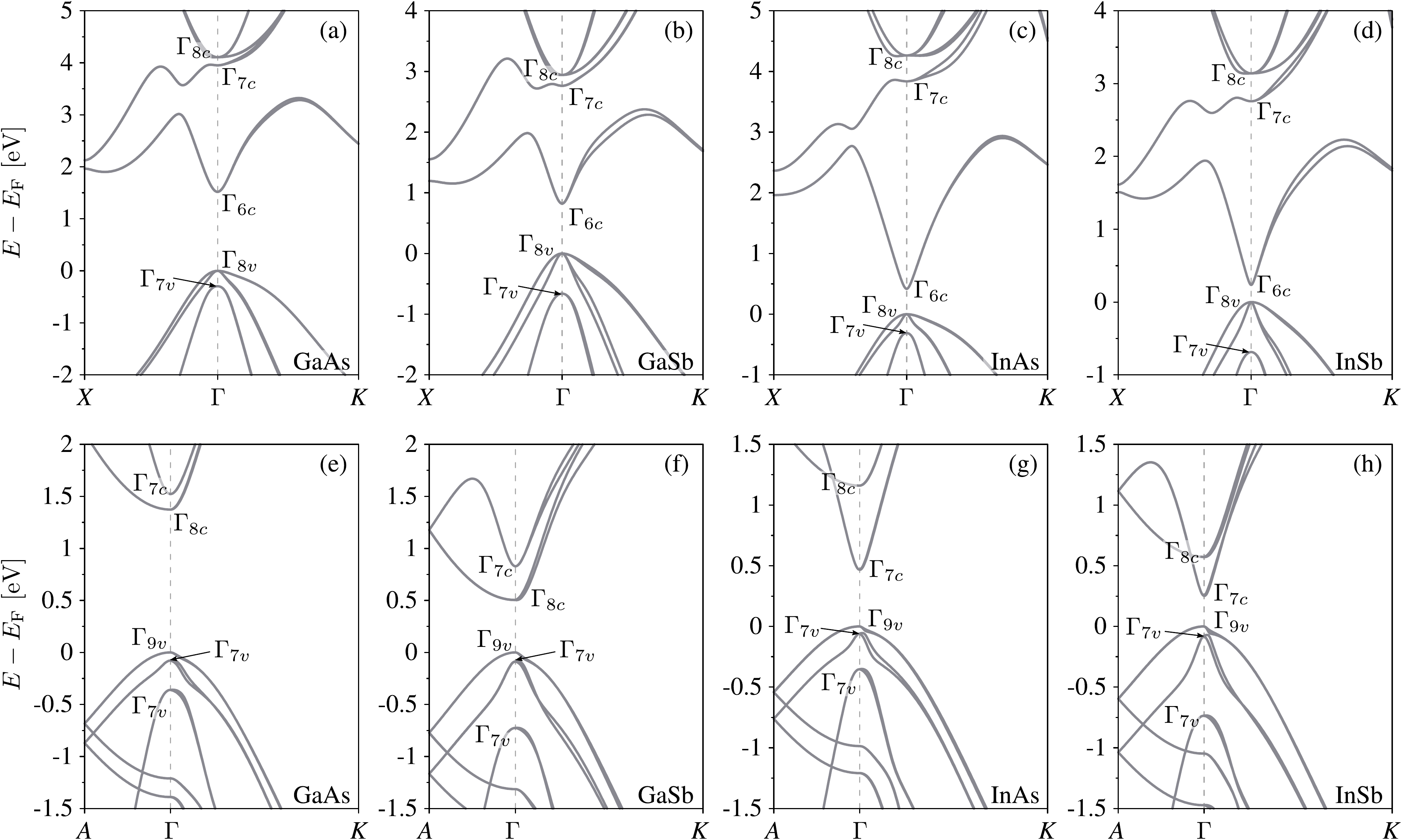}
	\caption{\label{Fig:bands}
Calculated band structures for III-V semiconductors. Zinc-blende phases for (a)~GaAs, (b)~GaSb, (c)~InAs, (d)~InSb, and wurtzite phases for (e)~GaAs,(f)~GaSb, (g)~InAs, (h)~InSb are shown along high symmetry lines crossing the zone center. The irreducible representations of the double symmetry groups at the zone center are also indicated,
according to Ref.~\onlinecite{Koster1963:book, Altmann1994:book}.
	}
\end{figure*}
%------------------------------------------------------------------

Here we calculate electronic band structures and the spin properties of selected III-V semiconductors using the full potential linearized augmented plane wave method as implemented in Wien2k code.\cite{wien2k} The wave functions are expanded in atomic spheres for orbital quantum numbers up to 10; the plane wave cut-off multiplied with the smallest atomic radii equals to 10. Relativistic local orbitals with $p_{1/2}$ radial wavefunctions are added for all elements to improve the basis set.\cite{Kunes2001:PRB} The irreducible Brillouin zone is sampled with 600 $k$ points. To overcome the afforementioned deficiencies of standard DFT calculations of the band gap underestimation, we use the modified Becke-Johnson exchange potential \cite{Becke2006:JCHP} plus LDA-correlation.\cite{Perdew1981:PRB}
%It has been demonstrated that such parametrization allows to calculate band gaps with an accuracy similar to very expensive GW calculations. \cite{Kim2010:PRB,Koller2011:PRB,Koller2012:PRB}
Finally, spin-orbit coupling is included in self-consistent calculations within second variational step.\cite{Singh2006:book}

Structural similarities between ZB and WZ phases suggest that the local electronic environment is also similar for the two phases, and therefore the crystal potentials will be nearly identical for WZ and ZB.\cite{De2010:PRB,Dacal2014:MRE} We use the mBJ exchange potential that reproduces experimental band gaps in the ZB phase to predict the electronic structure 
and spin properties of the WZ phase.

%-------------------------------------------------------------------
\section{Results}
%-------------------------------------------------------------------

The results below are obtained from DFT using the mBJ exchange potential combined with LDA for correlations potential \cite{Tran2009:PRL} including spin-orbit coupling within atomic spheres. 
%
%For the atomic sphere radii we considered in bohrs the following 2.30, 2.30 for Ga, As, respectively.
%grr to fit ZB gap and then use the same for WZ
%RMTs:
%GaAs: ZB & WZ: Ga 2.30, As 2.30
%GaSb: ZB & WZ: Ga 2.42, Sb 2.50
%InAs: ZB & WZ: In 2.50, As 2.40
%InSb: ZB & WZ: In 2.50, Sb 2.50
%
It has been shown in case of GaAs that electronic properties are sensitive to structural parameters.\cite{Cheiwchanchamnangij2011:PRB}
To provide realistic first-principles data we use the experimental lattice constants determined at low temperatures. Only in the case of wurtzite GaSb we lack reliable experimental data for the structure parameters, so we take ideal 
relations between the WZ and ZB lattice constant phases: $a_{\rm WZ}=a_{\rm ZB}/\sqrt{2}$, $c_{\rm WZ} = a_{\rm WZ} \sqrt{8/3}$, and $\epsilon=0$. 
We have scaled the mBJ potential to reproduce the experimental band gaps in the ZB phase, see Tab.~\ref{Tab:table} and use the same potential with the experimentally defined atomic structures (up to GaSb mentioned above) to obtain the electronic band structures for the WZ phases. In addition, we determine the internal cell parameter $\epsilon$ by minimizing the total energy; the obtained values are listed in Tab.~\ref{Tab:table}. 

The predictive capability of such an approach of taking for the lattice constants values measured in nanowires is very satisfactory, as we compare the calculated bandgaps to some known measured values from WZ nanowires.\cite{Furthmeier2014:APL,Heiss2011:PRB,Ketterer2011:PRB,Moller2012:NT} For instance, in InAs WZ nanowires the lower bound on the band gap was estimated to be 0.46~eV by means of optical emission using photoluminescence spectroscopy.\cite{Moller2012:NT} Our calculated value is 0.461 eV. For WZ GaAs nanowires the gap is experimentally estimated to be less than 1.52~eV.\cite{Furthmeier2014:APL,Heiss2011:PRB,Ketterer2011:PRB}. We get
1,38 eV.  (The bandgaps measured in nanowires can be somewhat larger than in bulk 
due to quantum confinement). We also capture the experimental trends\cite{De2010:PRB,Cheiwchanchamnangij2011:PRB,Heiss2011:PRB} that GaAs and GaSb in WZ phase have a smaller gap than in ZB, while InAs and InSb larger. 
We have also tested our results of the ZB phase of GaSb for the $L_{6v}-\Gamma_{6v}$ gap, for which we get 50~meV,  which also reasonably agrees with electroreflectance spectra measurements yielding 63~meV \cite{Joullie1981:PRB}. For $L_{6v}-\Gamma_{8v}$ we get 0.8722~eV, while the experiment 0.8922~eV.

%--- Kim 2010 PRB in abstract:
%However, the effective masses are generally overestimated by 20 – 30 \% using the mBJ local potential.\cite{Kim2010:PRB} We believe this to be related to incorrect nearest-neighbor hopping elements, which are little affected by the choice of the local potential

%-------------------------------------------------------------------
\subsection{Band structures}
%-------------------------------------------------------------------

Similarities in the crystal structures of the ZB and WZ phases translate to the similarity of their band structures via relations between the band gaps \cite{Yeh1994:PRB} and high symmetry points in their corresponding Brillouin zones.\cite{De2010:PRB} There are similar correspondences between the high symmetry directions of the two crystals as well. The symmetry line $\Lambda$ $(\Gamma \to L)$ in ZB corresponds to the $\Delta$ $(\Gamma \to A)$ line in WZ phase.\cite{Birman1959:PRL}

In Fig.~\ref{Fig:bands} we show the calculated band structures in the vicinity of the $\Gamma$ point for GaAs, GaSb, InAs, and InSb in ZB and WZ phases. All theses materials are central zone, direct band gap semiconductors. In Table~\ref{Tab:table} we list the band gaps $E_g$ obtained from experiment for the ZB which coincide with the scaled mBJ calculations and the calculated gaps for the WZ phases. The $T_d$ symmetry of the ZB reduces to the $C_{6v}$ symmetry in the WZ phase, giving rise to a crystal field which is manifested by an additional structure at the valence band edge. Without spin-orbit coupling, the top valence states in the ZB phase belong to the three dimensional $\Gamma_{5v}$ representation,  which is in the WZ phase further decomposed into the two dimensional $\Gamma_{5v}$ and one dimensional $\Gamma_{1v}$.

%---> The states at the top of the valence band in WZ also have some important differences with their ZB counterparts. In the absence of spin-orbit coupling, the hexagonal crystal field of WZ splits the p-like $\Gamma_{15}$ state of ZB into a fourfold degenerate $\Gamma_6$ and a doubly degenerate $\Gamma_1$. In terms of the p orbitals, these states are $p_z \to \Gamma_1$ are $p_x,p_y \to \Gamma_6$. 

%The top valence bands in the single group representation $\Gamma_{5v}$ in ZB phase are three-fold degenerated. Spin-orbit coupling lifts the degeneracy and splits the states to spin-orbit split-off band $\Gamma_{7v}$, in double group representation, and two-fold $\Gamma_{8v}$ for heavy holes (HH) and light holes (LH) branches by the energy of $\Delta_{\rm so}$. 

\begin{figure}[h!]
	\centering
	\includegraphics[width=0.98\columnwidth]{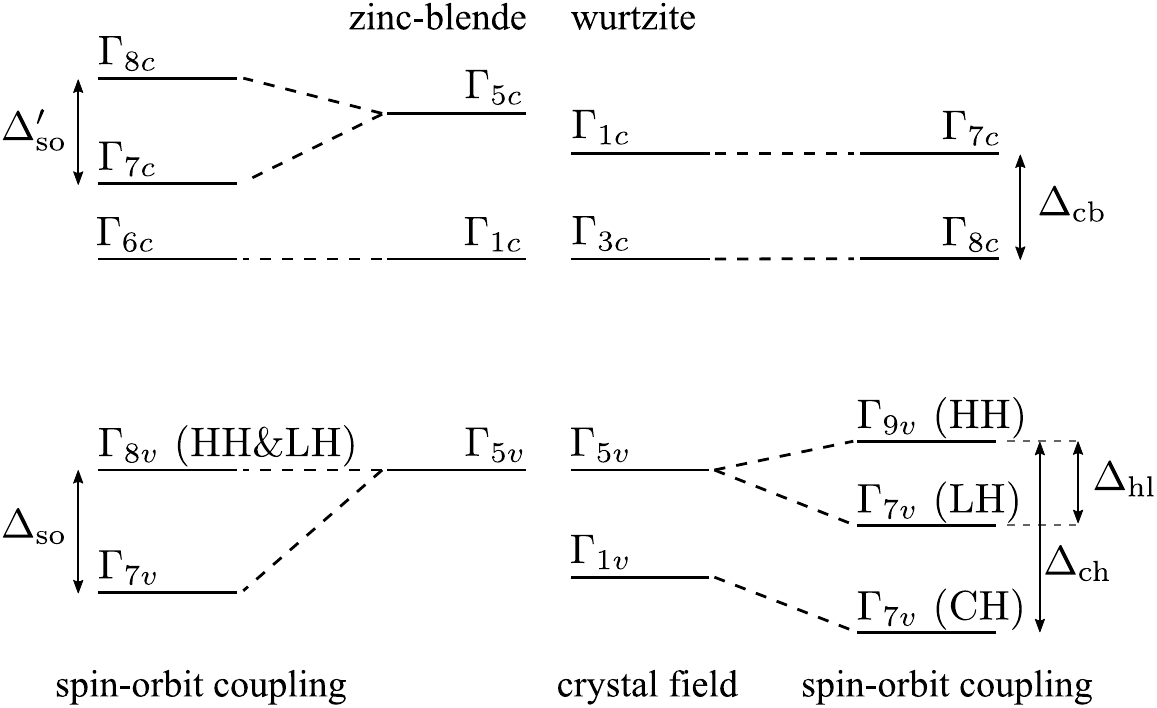}
	\caption{\label{Fig:scheme}
Scheme of the energy levels at the zone center in III-V zinc-blende and wurtzite semiconductors. The corresponding group symmetry representations and split-off energies between heavy holes (HH), light holes (LH), and crystal hole (CH) states as well as splitting of the conductance bands are listed in Tab.~\ref{Tab:table}.
	}
\end{figure}

Spin-orbit coupling further lifts the triple degeneracy of the valence bands of the $\Gamma_{5v}$ representation in the ZB phase to the four dimensional $\Gamma_{8v}$ double group representation and the spin split-off band in the two dimensional $\Gamma_{7v}$ representation. Calculated spin-orbit split-off energies $\Delta_{\rm so}$ are listed in Tab.~\ref{Tab:table}. The degenerate (at zone center) heavy holes (HH) and light-holes (LH) states belong to the $\Gamma_{8v}$ representation. In the WZ phase the reduced symmetry due to the crystal field leads to the splitting (at zone center) of the HH and LH by the energy $\Delta_{\rm hl}$, to two dimensional double group representations $\Gamma_{9v}$ and $\Gamma_{7v}$. The above discussion is summarized graphically 
in Fig.~\ref{Fig:scheme}.

%
%------------------------------------------------------------------
%------------------------------------------------------------------

The bottom of the conduction band edge in the ZB phase is formed by 
$\Gamma_{6c}$ states. Above the $\Gamma_{6c}$ there are spin-orbit split $\Gamma_{7c}$ and $\Gamma_{8c}$ bands; the calculated 
splitting is denoted as $\Delta'_{\rm so}$ in Tab.~\ref{Tab:table}. The 
energy difference between the conduction band edge of $\Gamma_{6c}$ representation and the valence band edge of $\Gamma_{8v}$ gives the band gap $E_g$. 

In the WZ phase the band gap is formed between the valence edge of $\Gamma_{9v}$ and the conduction band edge, which is $\Gamma_{8c}$ for GaAs and GaSb, and 
$\Gamma_{7c}$ InAs and InSb. The appearance of $\Gamma_{8c}$ states is a consequence of the zone folding due to the doubled unit cell of the WZ crystal along [111] direction. Therefore, an additional zone center transition is expected to appear coming from the $L$ point minimum in the ZB dispersion. The usual order of the conduction bands is $\Gamma_{7c}$, forming the conduction edge, followed by the $\Gamma_{8c}$,
which is only weakly coupled to the light.\cite{Peng2012:APL}
The calculated energy differences $\Delta_{\rm cb}$ between the $\Gamma_{8c}$ and $\Gamma_{7c}$ states at the zone center are listed in Tab.~\ref{Tab:table}.

Low temperature experiments show that GaAs in the WZ phase has a smaller gap \cite{Furtmeier2016,Kusch2012:PRB,Ketterer2011:PRB,Heiss2011:PRB} than in the ZB phase. This suggests that the minimum of the conduction band has indeed $\Gamma_{8c}$ representation, and not $\Gamma_{7c}$ in agreement with other calculations.\cite{De2010:PRB} We note that the experimental gaps are measured from the photoluminescence on GaAs WZ nanowires, which should have a slightly higher value 
than in the bulk, due to confinement. It has been also shown that the order of the representations in the conduction band edge is affected by strain.\cite{Cheiwchanchamnangij2011:PRB,Peng2013:PRB} A direct quantitative comparison of our calculated gaps in the WZ structures with the existing experiment is thus not yet
possible.

We also note that the calculated internal cell parameter $\epsilon$ in WZ GaAs, InAs and InSb is very small, less than 0.001, see Tab.~\ref{Tab:table}. This agrees well with other calculations\cite{Panse2011:PRB,Belabbes2012:PRB} and the experimental determination for InAs nanowires\cite{Zanolli2007:JOP}. On the other hand, this internal cell parameter for GaAs has been estimated to be two times larger\cite{McMahon2005:PRL} than 
our calculated value, but these measurements\cite{McMahon2005:PRL} were performed in a metastable bulk GaAs. Investigating the possible effects of  $\epsilon$ on spin-orbit coupling parameters, we have analyzed the electronic structures of WZ phase for $\epsilon=0$ and for the relaxed value of $\epsilon$, listed in Tab.~\ref{Tab:table}. We found that in the structures with the relaxed value of $\epsilon$ the mostly affected is InSb: 
$\Delta_{\rm hl}$ increases by 10\%.

%-------------------------------------------------------------------
\subsection{Spin-orbit coupling}
%-------------------------------------------------------------------

Describing semiconductor bandgaps with standard local and semilocal exchange-correlation functionals within DFT 
leads to an overestimation of the band spin splitting (due to the underestimation of the band gap), when one
compares with many-body treatments.\cite{Chantis2006:PRL,Luo2009:PRL} It has been shown that the magnitude of the spin-orbit coupling parameters depends on the hybridization between bands. \cite{Chantis2006:PRL}
In this section we present our results obtained using mBJ\cite{Becke2006:JCHP,Tran2009:PRL} with spin-orbit coupling treated in the second variational step.\cite{Singh2006:book} In known cases our results are in agreement with computationally more expensive many body approaches. We determine relevant spin-orbit coupling parameters by 
fitting symmetry-determined functional forms of the spin-orbit splittings to our DFT data close to the zone center.

Spin-orbit coupling splits orbital degeneracies of the electronic bands at high-symmetry points. In systems with
a space inversion symmetry, time reversal symmetry would lead to at least double degeneracy of the Bloch states. 
Neither ZB nor WZ phase of III-V semiconductors has space inversion symmetry, which is manifested 
by the spin-splitting of the states at a general $\mathbf{k}$ point.\cite{Dresselhaus1955:PR,Casella1959:PR}
Only time-reversal invariant points, for which $\mathbf{k}$ and $-\mathbf{k}$ are equivalent (differ by a reciprocal
lattice vector), the spin degeneracy is recovered. Trivially, $\Gamma$ point of the zone center is such a point. Other
high symmetry time-reversal invariant points are lattice specific.
In general, we can conclude on the existence of spin splitting by analyzing the dimensionality of the irreducible representations of the double groups
\cite{Elliot1954:PR,Koster1963:book,Altmann1994:book} at high symmetry points and lines.

%-------------------------------------------------------------------
\subsubsection{Zinc-blende structures}
%-------------------------------------------------------------------
%

\begin{figure}[h!]
	\centering
	\includegraphics[width=0.99\columnwidth]{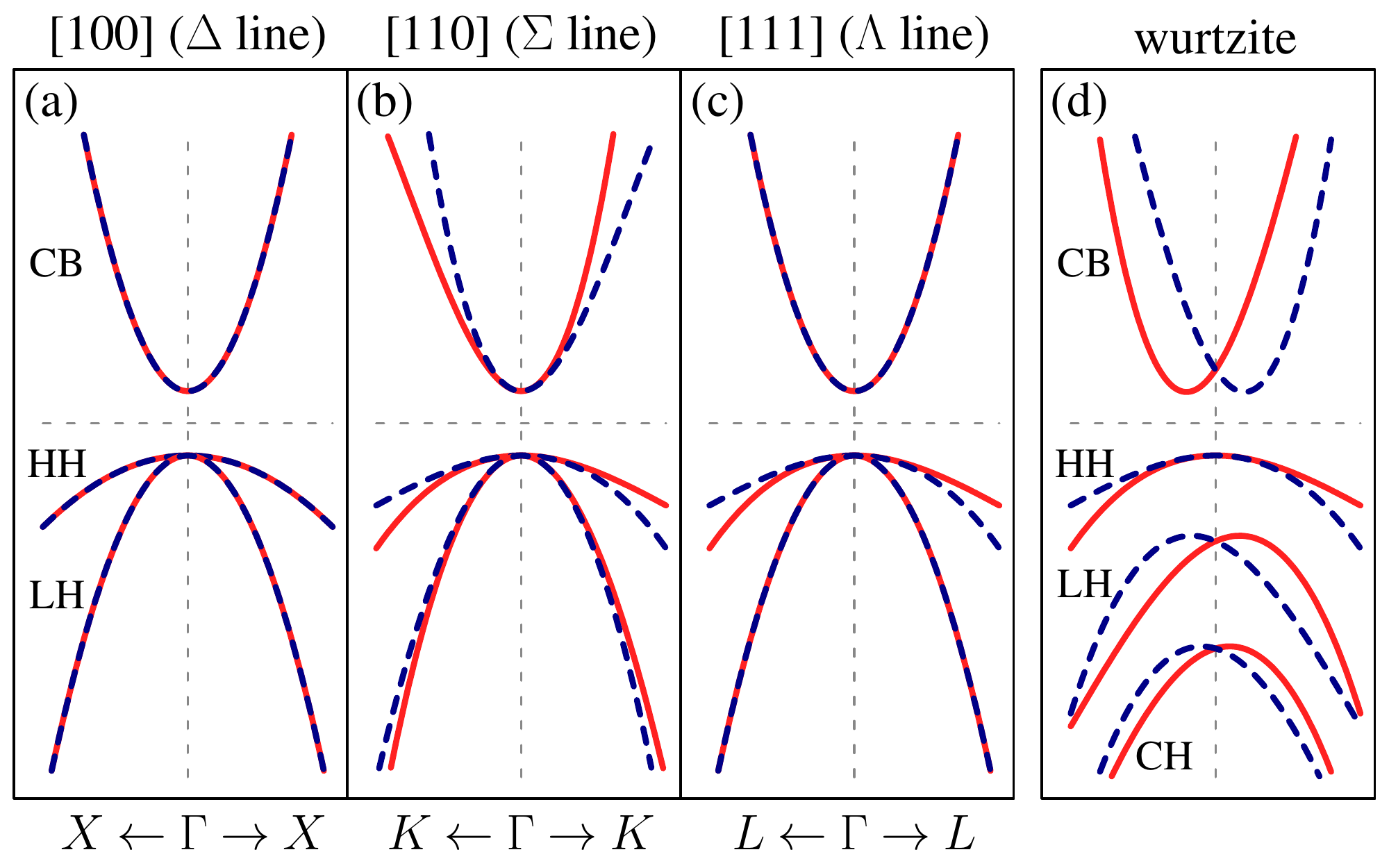}
	\caption{\label{Fig:bands_splitting}
		Schematic plot of bands spin splitting near the zone center for spin up and spin down shown with solid and dashed curves. The splitting is shown for zinc-blende along directions (a)~[100] the $\Delta$ line; (b)~[110] the $\Sigma$ line; (c)~[111] the $\Lambda$ line; and for (d)~wurtzite structure along any direction except the $\Delta$ line. The symbol CB denotes conduction bands, HH stands for heavy holes, LH for light holes and CH for crystal hole bands.
	}
\end{figure}

The time-reversal invariant points of ZB structures are $\Gamma$, $L$, and $X$. At these points the bands are always doubly spin degenerate. In addition, the Bloch states along [100], i. e.,  along the $\Delta$ line, have the $C_{2v}$ symmetry whose double group has only two dimensional irreducible representations. Therefore, the states along $\Delta$ do not spin-split either, see Fig.~\ref{Fig:bands_splitting}(a). 
%------------------------------------------------------------------
%------------------------------------------------------------------

Along the [110] direction, the $\Sigma$ line, including the $K$ points, the states transform by the $C_s$ point group. Since $C_s$ has only one dimensional double group representations, all the states are in general spin-split, see Fig.~\ref{Fig:bands_splitting}(b), to $\Sigma_3$ or $\Sigma_4$ representations. The splitting close to zone center is proportional to $k^3$ for bands with $\Gamma_{6}$ and $\Gamma_{7}$ representations. However splitting of the $\Gamma_{8}$ contains also terms linear in $k$ of the form $\Delta E = C k+\gamma k^3$, where $C$ and $\gamma$ are the corresponding energy splitting parameters.\cite{Dresselhaus1955:PR,Cardona1986:PRL,Cardona1988:PRB} We determine the spin-orbit coupling parameters by fitting the $\Delta E$ in the vicinity of the $\Gamma$ point up to 2\% of the $\Gamma - K$ width to our calculated DFT data. The signs of the spin-orbit coupling parameters depend on the specific atomic positions (ordering of cations and anions) within the unit cell\cite{Cardona1986:PRL,Cardona1988:PRB} and orientation of the quantization axis. To uniquely determine the spin split states we consider the $\Sigma$ line that points towards $K$ point with $[3/4,3/4,0]$ coordinates in conventional basis and spin quantization along [001]. In this case the reflection of the $C_s$ point group, the plane (110), multiplies state spin-up $(\Sigma\uparrow)$ with $-i$ while spin-down state $(\Sigma\downarrow)$ by $i$, and hence $(\Sigma\uparrow)$ belongs to $\Sigma_4$ and $(\Sigma\downarrow)$ to $\Sigma_3$ representation according to the character table in Ref.~[\onlinecite{Koster1963:book}]. We define the energy spin splitting $\Delta E$ as positive if the spin-up state is above the spin-down state.
%state with $\Sigma_4$ is above the $\Sigma_3$ state or the spin-up state $(\Sigma\uparrow)$ is above the $(\Sigma\downarrow)$ state.

States along the [111] direction, the $\Lambda$ symmetry line, belong to the $C_{3v}$ point group and may fall to one or two dimensional irreducible double group representations. Therefore, the states along  $\Lambda$ may, but need not spin split. For our materials the conduction and light-hole bands along $\Lambda$ line do not split, while the valence bands of $\Gamma_{8v}$ symmetry, heavy holes, do spin-split to $\Lambda_5$ or $\Lambda_6$ representations, also following the $C k+\gamma k^3$ dependence close to the zone center.\cite{Dresselhaus1955:PR,Cardona1986:PRL,Cardona1988:PRB} See Fig.~\ref{Fig:bands_splitting}(c).
To determine the sign of the $C$ and $\gamma$ parameters, we consider $L$ point $[1/2,1/2,1/2]$ in conventional basis and spin quantization axis along [001]. The reflection plane $\sigma_v$ distinguishes spin-up and spin-down states that belong to $\Lambda_6$ and $\Lambda_5$ representations.\cite{Koster1963:book} Similarly, the spin splitting $\Delta E$ is defined as positive when the spin-up state is above the spin-down state, or when states $\Lambda_6$ are above $\Lambda_5$ states.\cite{Cardona1988:PRB} 
Comparing especially the cubic parameters for the HH band along $\Sigma$ and $\Lambda$ lines, see Tab.~\ref{Tab:table}, we find them strongly anisotropic. We observe also that the absolute values of the spin-orbit parameters for the studied semiconductors in general grow with the atomic weight of the compounds. However, the parameters describing linear in $k$ spin-splittings the atomic weight of the cation plays the dominant role.

In Table~\ref{Tab:table} we show the calculated spin-orbit coupling parameters for the valence bands as well as the split-off gap at zone center, $\Delta_{\mathrm{so}}$, for the valence (difference between $\Gamma_{8v}$ and $\Gamma_{7v}$), and $\Delta'_{\mathrm{so}}$, for the conduction (difference between $\Gamma_{8c}$ and $\Gamma_{7c}$) bands; see Fig.~\ref{Fig:scheme}.
These gaps reflect on the strength of the spin-orbit coupling at the anion and cation sites, as the zone center coupling is directly related to the atomic fine structure of the atoms supplying the principal band orbitals. We find that $\Delta_\mathrm{so}$ is mainly controlled by the the anion (As, Sb), while $\Delta_\mathrm{so}'$ by the cation (Ga,In).\cite{Chantis2006:PRL} 

\begin{figure}[h!]
	\centering
	\includegraphics[width=0.98\columnwidth]{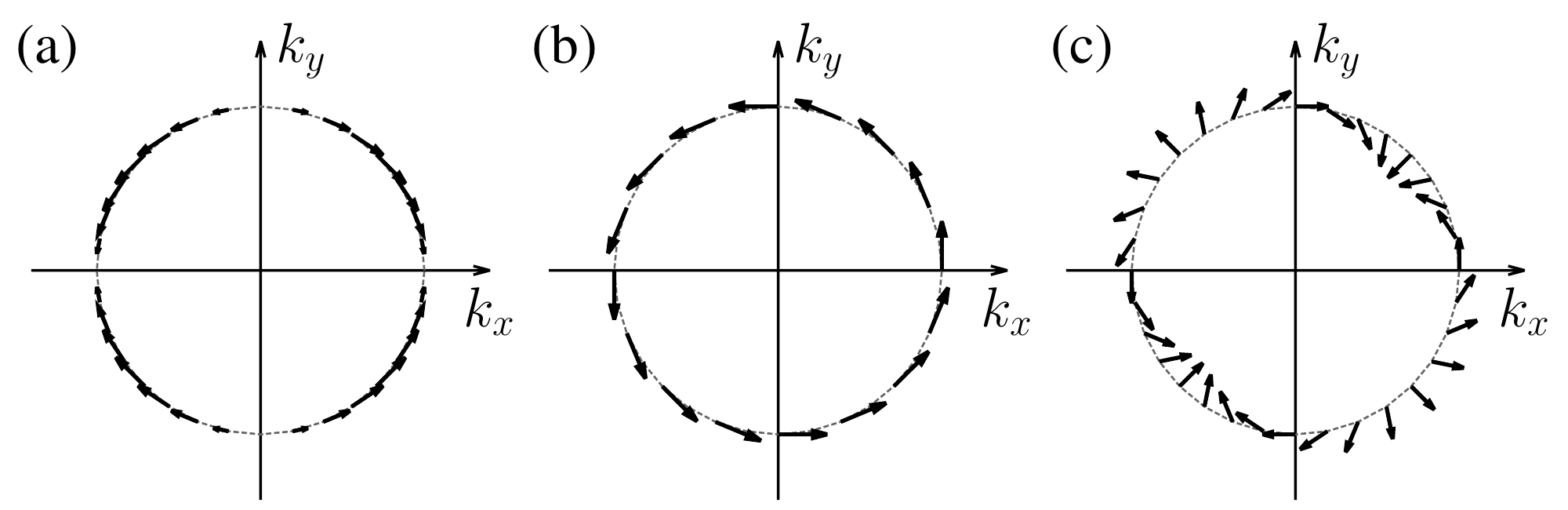}
	\caption{\label{Fig:SOCF}
		Contour plots of spin-orbit fields $\mathbf{\Omega}(\mathbf{k})$ near the zone center in III-V semiconductors for $k_z=0$. 
		(a)~Vector field for the conduction states with $\Gamma_6$ symmetry in the ZB phase.
		(b)~Same in the WZ phase for states with  $\Gamma_7$ and $\Gamma_8$ symmetries, and
		(c)~for  $\Gamma_9$ symmetry.
	}
\end{figure}

\begin{figure}[h!]
	\centering
	\includegraphics[width=0.98\columnwidth]{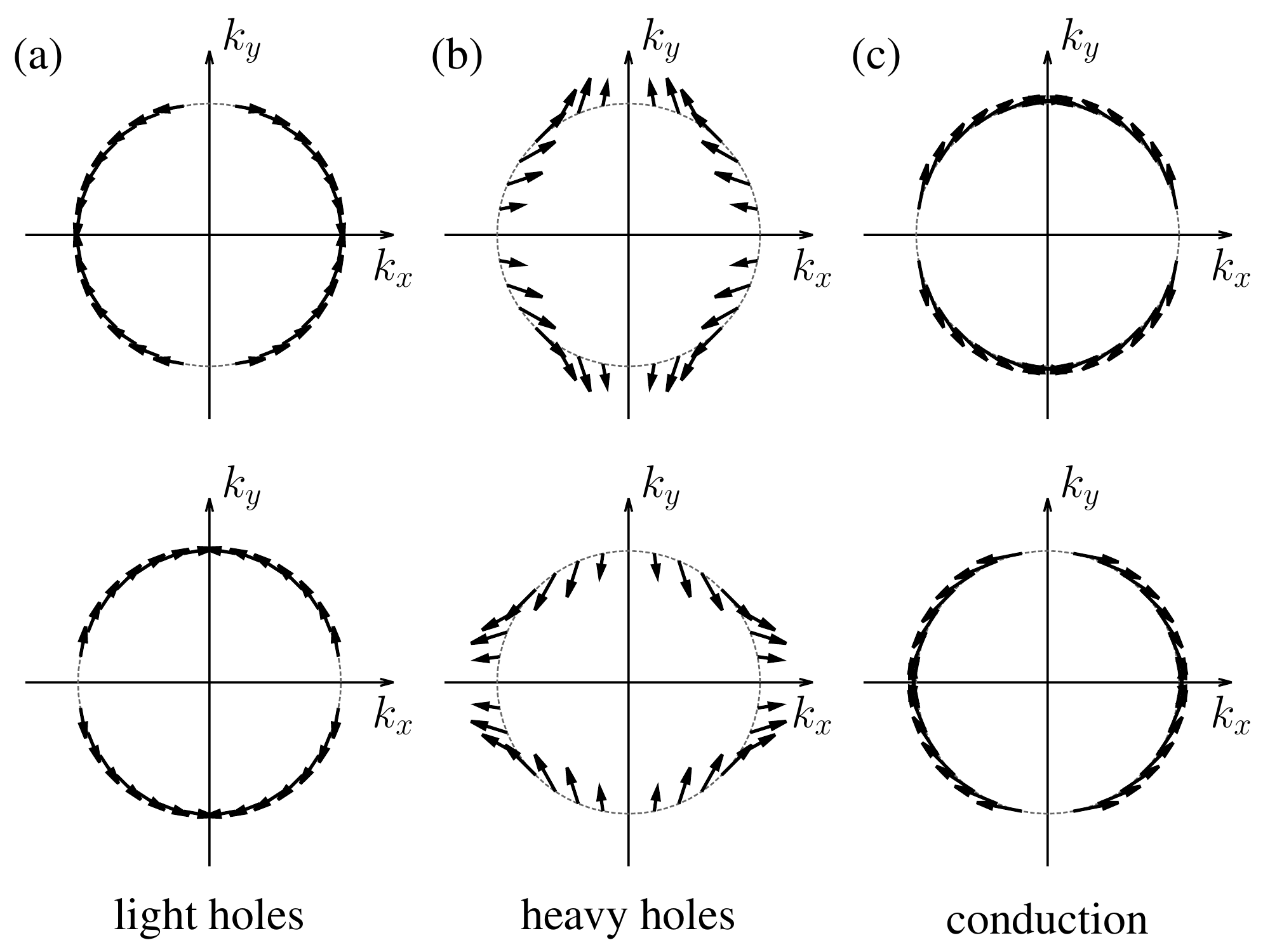}
	\caption{\label{Fig:spins_ZB}
		Calculated spin expectation values for zinc-blende GaAs. The momentum contour around zone center with $k$ equal to 1\% of Brillouin zone width and $k_z=0$. (a)~spins for spin split light hole bands, (b)~for heavy hole bands, and (c)~for conduction bands. The bottom row corresponds to the bands (of the spin-split family) with the lower energy.
	}
\end{figure}

The spin splitting near the $\Gamma_{6c}$ conduction band minimum can be described by the operator 
\begin{equation}
H_{\rm sof} = \frac{\hbar}{2}\boldsymbol{\sigma}\cdot\mathbf{\Omega}(\mathbf{k}).
\end{equation}
Here $\boldsymbol{\sigma}$ is the vector of Pauli spin matrices and $\mathbf{\Omega}(\mathbf{k})$ is the spin-orbit field (labeled as sof). For the ZB conduction bands the functional form of this field  is 
\begin{equation}
\mathbf{\Omega}(\mathbf{k})=\gamma [k_x(k_y^2-k_z^2),k_y(k_z^2-k_x^2),k_z(k_x^2-k_y^2)],
\end{equation}
as first derived by Dresselhaus.\cite{Dresselhaus1955:PR} In Fig.~\ref{Fig:SOCF}(a) we plot this cubic
Dresselhaus spin-orbit field as a vector field on a momentum contour for $k_z=0$. As the conduction band is formed from $s$-type orbitals, their spin would point along the $\mathbf{\Omega}(\mathbf{k})$; compare to 
the calculated spin expectation values shown in Fig.~\ref{Fig:spins_ZB}(c). The hole states are formed by the $p$-type orbitals which are split due to SOC to the total angular momentum $J=3/2$ and $J=1/2$ separated by the $\Delta_{\rm so}$ energy. In this case the SOC field can not be directly expressed, although the spin expectation values of the Bloch states can be calculated, what we show  in Fig.~\ref{Fig:spins_ZB}(a,b).
%
%------------------------------------------------------------------
%------------------------------------------------------------------

The field $\mathbf{\Omega}(\mathbf{k})$ is responsible for conduction band $\Gamma_{6c}$ spin splitting which for small $k$ varies as $k^3$. The proportionality parameter $\gamma$ depends on the bulk properties of the material and its amplitude often grows with anion mass and scales as $1/E_g$ for narrow gap semiconductors.\cite{Chantis2006:PRL} We determine the sign of $\gamma$ by fixing $\mathbf{\Omega}(\mathbf{k})$ to the calculated spin expectation values, compare Fig.~\ref{Fig:SOCF}(a) and Fig.~\ref{Fig:spins_ZB}(c) for GaAs. We note that the sign of $\gamma$ can be also determined from symmetry representations of the eigenvectors near the zone center.\cite{Cardona1988:PRB,Chantis2006:PRL} Calculated spin expectation values for GaSb, InAs, and InSb are similar to the GaAs case shown in Fig.~\ref{Fig:spins_ZB}, see Appendix.

We determine the $\gamma$ parameter for the $\Gamma_{6c}$ band in GaAs to be 9.13~${\rm eV\AA^3}$ which is in good agreement with previous calculations. For instance, empirical pseudopotentials corrected by fitting to GW\cite{Luo2009:PRL} found the corrected value of 8.3~${\rm eV\AA^3}$ with respect to the LDA value of 46.8~${\rm eV\AA^3}$. Adding empirical pseudo Darwin potential shifts\cite{Cardona1988:PRB} to adjust gaps at high symmetry points to reproduce experimental gaps results in a qualitatively reasonable band structure with $\gamma$ calculated within LMTO to 14.9~${\rm eV\AA^3}$ and $k\cdot p$ theory to 29.8~${\rm eV\AA^3}$. A semiclassical billiard model \cite{Krich2007:PRL} provide 9~${\rm eV\AA^3}$. Extensive self-consistent GW calculations with the spin-orbit interaction taken as a perturbation to the scalar relativistic Hamiltonian and scaled self-energy to reproduce experimental bandgaps\cite{Chantis2006:PRL} give 8.5~${\rm eV\AA^3}$. Comparing our calculated $\gamma$ parameter for GaSb, InAs, and InSb with the GW calculations \cite{Chantis2006:PRL}, see Tab.~\ref{Tab:table}, we conclude that the employed mBJ potential gives accurate results.

On a qualitative level, $\gamma$ for the $\Gamma_{6c}$ conduction band grows with the total atomic weight of the compound. For the valence states the amplitude of the cubic spin-orbit coupling parameter is controlled by the anion type while the linear parameters by the cation.

%------------------------------------------------------------------
%------------------------------------------------------------------

%-------------------------------------------------------------------
\subsubsection{Wurtzite structures}
%-------------------------------------------------------------------

Time reversal invariant points of the WZ structure are $\Gamma$, $M$, and $A$. Here the spin-splitting is absent. Also, the $\Delta$ line connecting $\Gamma$ and $A$ has the $C_{6v}$ symmetry whose double group representations
are all two dimensional, so also along this line the spin-orbit fields vanish. 
%It has been pointed out that this two-fold spin degeneracy will exist for all energy bands in crystals of the WZ structure.\cite{Casella1959:PR,Hopfield1961:JAP}
At all other points in the Brillouin zone we expect spin splitting, except for accidental degeneracies. The spin splitting of the bands close to the zone center is schematically shown in Fig.~\ref{Fig:bands_splitting}(d). There are no special directions along which the spin-orbit fields vanish, in contrast to the ZB structure, except for the mentioned $\Delta$ line.

We have extracted the energy differences $\Delta_\mathrm{cb}$ between the first and second conduction bands in the
zone center. The positive value is for $\Gamma_{8c}$ states higher in energy than $\Gamma_{7c}$. The calculated $\Delta_\mathrm{cb}$ are in Tab.~\ref{Tab:table}. The sign is determined by the cation element. For Ga $\Delta_\mathrm{cb}$ is negative, while for In it is positive. On the other hand, the crystal field offset of the $\Gamma_{7v}$ band from the top of the valence band, $\Delta_\mathrm{ch}$, is controlled by the anion type and is almost independent on the cations. Spin-orbit coupling splittings of the HH and LH, $\Delta_\mathrm{hl}$ are found in the range from 60 to 90 meV. 
%In the literature one often founds splitting of the valence band edge in terms of the spin-orbit parameter $\Delta_{\rm so}$ and the crystal field parameter $\Delta_{\rm cr}$ derived from $\mathbf{k}\cdot\mathbf{p}$ theory\cite{Chuang1996:PRB}, $E(\Gamma_{9v})-E(\Gamma_{7v}^{\rm LH/CH}) = (\Delta_{\rm cr}+\Delta_{\rm so}\pm\sqrt{(\Delta_{\rm cr}+\Delta_{\rm so})^2-8\Delta_{\rm cr}\Delta_{\rm so}/3})/2$.
Calculated energy splittings are collected in Tab.~\ref{Tab:table}.

The absence of inversion symmetry in the WZ structure allows also terms linear in $k$ in the electronic band structure when the spin-orbit interaction is included.\cite{Rashba1960:SPSS,LewYanVoon1996:PRB} The possible presence of the linear terms was a hot debated topic early on in the investigations of WZ semiconductors, both in theory \cite{Casella1959:PR,Hopfield1961:JAP,Mahan1964:PR,LewYanVoon1996:PRB} and experiment. \cite{Hummer1978:PSSB,Broser1979:PSSB,Koteles1980:PRL} It was shown later that the linear spin-splittings are very sensitive to the accuracy of the band gap determination.\cite{LewYanVoon1996:PRB} 
This suggests that the overall value for the spin-orbit coupling in typical WZ semiconductors depends on the precise position of the bands and their hybridization.

Using group theory and $k\cdot p$ modeling analytical expressions for the spin-orbit coupling fields of electrons and holes due to the bulk inversion asymmetry in WZ semiconductors have been derived.\cite{Fu2008:JAP,Wang2007:APL} The functional form of the spin-orbit field for $\Gamma_7$ and $\Gamma_8$ states close to the zone center is
\begin{equation}
 \mathbf{\Omega}(\mathbf{k})=(\alpha + \gamma[b k_z^2-k_\parallel^2])(k_y,-k_x,0).
\end{equation}
 For the $\Gamma_9$ states the spin-orbit field depends on the momentum as
\begin{equation}
\mathbf{\Omega}(\mathbf{k})=\gamma [k_y (k_y^2 - 3 k_x^2), k_x (k_x^2-3k_y^2),0].
\end{equation}
Both vector fields are plotted schematically in Fig.~\ref{Fig:SOCF}(b,c). One notes that the components of the spin-orbit fields $\mathbf{\Omega}(\mathbf{k})$ in the WZ phase have in-plane components only, perpendicular to the hexagonal axis. 

Bands with $\Gamma_7$ and $\Gamma_8$ symmetries consist of both cubic and linear terms in $k$.\cite{Fu2008:JAP} The linear term in $k$ originates from the $C_{6v}$ point group symmetry of the WZ phase.\cite{Rashba1960:SPSS}
It leads to a linear energy splitting close to the zone center. The spin splitting is proportional to the parameter $\alpha$. For larger $k$ the splitting grow as $k^3$ and is proportional to parameters $\gamma$ and $b$. Parameter $b$ relates the splitting with the $k_z$ momentum, parallel to the hexagonal axis. We found that its value for valence bands is sensitive to the cell parameter $\epsilon$. For instance, if we take $\epsilon=0$,
$b$ in this case for GaAs and CH band is $-0.88$ (compared to -0.03 for relaxed
structure), LH $-0.09$ (versus -0.02), and for LH in InAs is equal to $-0.41$ (versus 0.49). Except for these, the influence of $\epsilon$ is less than 10\%, indeed negligible when 
considering the current experimental accuracy of determining these spin-orbit parameters.

The linear term of $\mathbf{\Omega}(\mathbf{k})$ for $\Gamma_7$ and $\Gamma_8$ states manifests in the band spin splitting of the same form as the Rashba splitting\cite{Rashba1960:SPSS}, see Fig.~\ref{Fig:bands_splitting}(d). The $\Gamma_{9v}$ states of the heavy holes in WZ do not have linear spin-orbit fields \cite{Casella1959:PR,Mahan1964:PR}, similarly to the zinc-blende $\Gamma_{6c}$ spin splitting shown in Fig.~\ref{Fig:bands_splitting}(b). However, in contrast to ZB, the WZ phase, both linear $\alpha$ and cubic $\gamma$ parameters, depend, for the conduction band $\Gamma_{8c}$, on the type of the anion. For $\Gamma_{7c}$ states as well as for the hole bands $\Gamma_{9v}$ and $\Gamma_{7v}$, the overall atomic weight of the compound gives the strength of spin-orbit coupling.

%-----------------------------------------------------------------------------------------------------------
\begin{table*}
\caption{Structural and spin-orbit coupling parameters of studied III-V semiconductors in zinc-blende and wurtzite phase. Calculated bandgaps coincide with the experimental values for zinc-blende structures with a given accuracy. Experimentally determined values are indicated by giving the corresponding references. The values in parantheses are 
GW calculations with scaled self-energy to reproduce experimental bandgaps. }\label{Tab:table}
\begin{ruledtabular}
\begin{tabular}{llcccc}
 Parameter & & GaAs & GaSb & InAs & InSb \\
\hline
\\
\multicolumn{6}{l}{zinc-blende}\\
\cline{1-2}\\
a [\AA]          &  & 5.65325\footnotemark[1] & 6.09588\footnotemark[2] & 6.0583\footnotemark[1] & 6.4794\footnotemark[1]\\
$E_g$ [eV]       &  & 1.519\footnotemark[1]   & 0.822\footnotemark[2]   & 0.417\footnotemark[1]  & 0.235\footnotemark[1] \\
\\
\multicolumn{2}{l}{$\Delta_\mathrm{so}$  [eV] [$E(\Gamma_{8v})-E(\Gamma_{7v})$]} & 0.294 & 0.66  & 0.31  & 0.685 \\
 exper.       &                              & 0.346\footnotemark[2] & 0.756\footnotemark[2] & 0.38\footnotemark[2] & 0.81\footnotemark[2] \\
\multicolumn{2}{l}{$\Delta_\mathrm{so}'$ [eV] [$E(\Gamma_{8c})-E(\Gamma_{7c})$]} & 0.156 & 0.178 & 0.426 & 0.383 \\
 exper.       &                              & 0.171\footnotemark[2] & 0.213\footnotemark[2] & -- & 0.39\footnotemark[2] \\
\\
$\Gamma_{6c}$ & \multicolumn{5}{l}{$\mathbf{\Omega}(\mathbf{k})=\gamma [k_x(k_y^2-k_z^2),k_y(k_z^2-k_x^2),k_z(k_x^2-k_y^2)]$}\\
              & $\gamma$ [${\rm eV\,\AA^3}$] &  9.13 &  105.3 &  -21.4 & 200 \\
              & (GW calc.)                   & (8.5)\footnotemark[3] &  (119.3)\footnotemark[3] & (-47.5)\footnotemark[3] & (209.6)\footnotemark[3] \\
\\
\multicolumn{2}{l}{spin-orbit coupling parameters for $\mathbf{k} || [110]$ ($\Sigma$ symmetry line) } \\
$\Gamma_{8v}$ & \multicolumn{5}{l}{$\Delta E(k)=C k + \gamma k^3$} \\
         (HH) & $C$      [${\rm meV\,\AA}$]  &  12.7 &   3.98 &  43.5 & 32.8 \\
              & $\gamma$ [${\rm eV\,\AA^3}$] & -3.33 & -13.85 &     2 & -7.14 \\
         (LH) & $C$      [${\rm meV\,\AA}$]  &  3.85 &   1.43 &  13.8 & 11.16 \\
              & $\gamma$ [${\rm eV\,\AA^3}$] &  31.3 &   160  & -43.7 & 216 \\
\\
\multicolumn{2}{l}{spin-orbit coupling parameters for $\mathbf{k} || [111]$ ($\Lambda$ symmetry line) } \\
$\Gamma_{8v}$ (HH) & $C$      [${\rm meV\,\AA}$]   &  13.5 &  4.31 &   46.9 &  35.46 \\
                   & $\gamma$ [${\rm meV\,\AA^3}$] & -53.6 & -53.7 & -179.4 & -209.5 \\

\\
%--------------------------------------------------------------------------------------------------------------------------------
\multicolumn{6}{l}{wurtzite} \\
\cline{1-2}\\
a [\AA]            & & 3.989\footnotemark[4] & 4.310 & 4.2742\footnotemark[5] & 4.5712\footnotemark[5] \\
c [\AA]            & & 6.564\footnotemark[4] & 8.145 & 7.025\footnotemark[5]  & 7.5221\footnotemark[5] \\
$\epsilon$         & & -0.00086 & -- & -0.00078 & -0.00097 \\
$E_g$ [eV]         & & 1.378 & 0.503 & 0.461 & 0.242 \\
%(experiment)       & & (1.46)\footnotemark[5]  &  &                        &       \\
\\
\multicolumn{2}{l}{$\Delta_\mathrm{cb}$ [eV] [$E(\Gamma_{8c})-E(\Gamma_{7c})$]}                                   & -0.135 & -0.323 & 0.706 & 0.337 \\
\multicolumn{2}{l}{$\Delta_\mathrm{hl}$ [eV] [$E(\Gamma_{9v})-E(\Gamma_{7v})$; splitting of the $\Gamma_{5v}$]}   &  0.089 &  0.087 & 0.066 & 0.091 \\
\multicolumn{2}{l}{$\Delta_\mathrm{ch}$ [eV] [$E(\Gamma_{9v})-E(\Gamma_{7v})$; crystal-field band offset]}        &  0.376 &  0.725 & 0.360 & 0.745 \\
%\multicolumn{2}{l}{$\Delta_\mathrm{so}^\mathrm{cr}$ [eV] (shift of the $\Gamma_{7v}$ w.r.t. $\Gamma_{1v}$)} &  &  &  &  \\
\\
$\Gamma_{7c}  $ &\multicolumn{5}{l}{$\mathbf{\Omega}(\mathbf{k})=(\alpha + \gamma[b k_z^2-k_\parallel^2])(k_y,-k_x,0)$}\\
& $\alpha$  [${\rm eV\,\AA}$]   & 0.04 & 0.078 & 0.3   & 0.71 \\ 
& $\gamma$  [${\rm eV\,\AA^3}$] & 6.51 & 52.1  & 132.5 & 892  \\ 
& $b$                           & 0.53 & 1.29  & -1.24 & -0.91 \\
$\Gamma_{8c}$ & \multicolumn{5}{l}{$\mathbf{\Omega}(\mathbf{k})=(\alpha + \gamma[b k_z^2-k_\parallel^2])(k_y,-k_x,0)$}\\
& $\alpha$  [${\rm eV\,\AA}$]   & 0.1  & 0.49  & 0.04  & 0.34 \\ 
& $\gamma$  [${\rm eV\,\AA^3}$] & 1.92 & 18.7  & 2.73  & 10.7 \\ 
& $b$                           & 0.06 & -0.04 & -0.06 & -0.07 \\
$\Gamma_{9v}$ (HH) & \multicolumn{5}{l}{$\mathbf{\Omega}(\mathbf{k})=\gamma (k_y (k_y^2 - 3 k_x^2), k_x (k_x^2-3k_y^2),0)$}\\
& $\gamma$  [${\rm eV\,\AA^3}$] & 69 & 187 & 521 & 1541 \\ 
%LH signs for alpha & gamma reversed according to spin orientation of the LH band
$\Gamma_{7v}$ (LH) & \multicolumn{5}{l}{$\mathbf{\Omega}(\mathbf{k})=(\alpha + \gamma[b k_z^2-k_\parallel^2])(k_y,-k_x,0)$}\\
& $\alpha$  [${\rm eV\,\AA}$]   & -0.36 & -0.44  & -0.74  & -1.15 \\ 
& $\gamma$  [${\rm eV\,\AA^3}$] &  71   &  123.6 &  420.8 &  608 \\ 
& $b$                           & -0.02 & -0.93  &  0.49  & 1.14 \\ 
$\Gamma_{7v}$ (CH) & \multicolumn{5}{l}{$\mathbf{\Omega}(\mathbf{k})=(\alpha + \gamma[b k_z^2-k_\parallel^2])(k_y,-k_x,0)$}\\
& $\alpha$  [${\rm eV\,\AA}$]   & 0.29  & 0.27   & 0.42 & 0.38 \\ 
& $\gamma$  [${\rm eV\,\AA^3}$] & -8.24 & 17     & -26  & 16.6 \\ 
& $b$                           & -0.03 & 3.8    &  2.6 & 1.8 \\ 
\end{tabular}
\footnotetext[1]{Ref.~\cite{Vurgaftman2001:JAP}}
\footnotetext[2]{Ref.~\cite{Madelung2004:book,Bublik1982:pssa,Nilsen2010:JVST}}
\footnotetext[3]{Ref.~\cite{Chantis2006:PRL}}
\footnotetext[4]{Ref.~\cite{McMahon2005:PRL}}
\footnotetext[5]{Ref.~\cite{Kriegner2011:NL}}
\footnotetext[6]{Ref.~\cite{Kusch2012:PRB}}
\end{ruledtabular}
\end{table*}
%-----------------------------------------------------------------------------------------------------------

%-------------------------------------------------------------------
\section{Conclusion}
%-------------------------------------------------------------------
We have studied the electronic structures, and in particular the spin splitting in ZB and WZ phases of GaAs, GaSb, 
InAs, and InSb semiconductors, by means of semilocal mBJ exchange potential within the DFT framework. 
We have found that this method gives accurate results, as judged by comparing with the GW. Indeed, 
the calculated spin-orbit coupling parameters for the zinc-blende phases agree well with known quasiparticle many body calculations, giving a strong support for the predictive power of our resutls for the wurtzite phases. 
We believe that this approach can be used to investigate spin-orbit coupling effects at semiconductor 
interfaces and surfaces, but also as a starting point (benchmark for fitting the band structure) for empirical methods.

%-------------------------------------------------------------------
\section*{Acknowledgments}
%-------------------------------------------------------------------
 We acknowledge support by DFG SFB 689.

%-------------------------------------------------------------------
\appendix
\section*{Appendix}
\begin{figure}[h!]
	\centering
	\includegraphics[width=0.98\columnwidth]{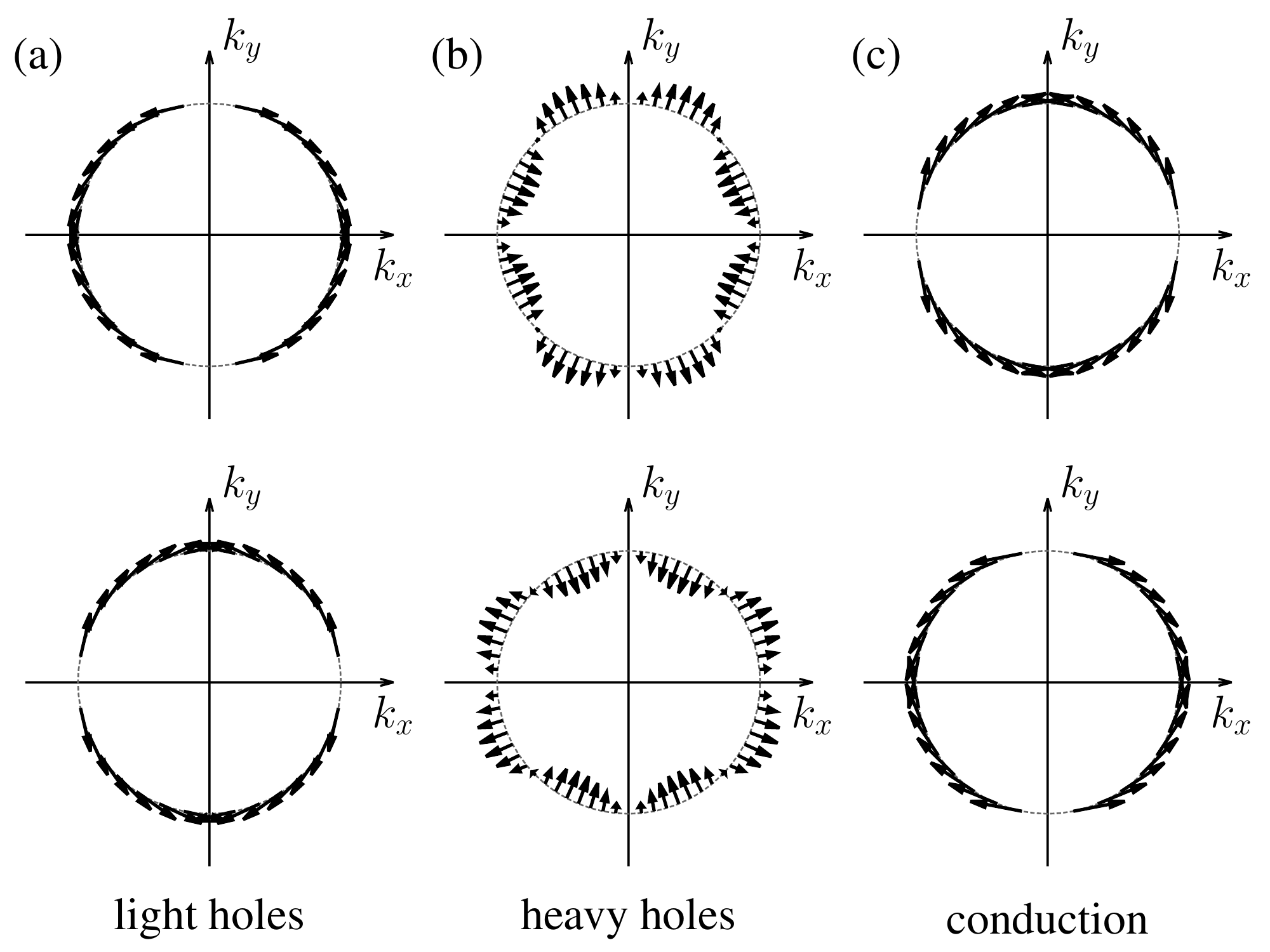}
	\caption{\label{Fig:App_spin_GaSb}
		Calculated spin expectation values for zinc-blende GaSb. The momentum contour around zone center with $k$ equal to 1\% of Brillouin zone width and $k_z=0$. (a)~spins for spin split light hole bands, (b)~for heavy hole bands, and (c)~for conduction bands. The bottom row corresponds to the bands (of the spin-split family) with the lower energy.
	}
\end{figure}

\begin{figure}[h!]
	\centering
	\includegraphics[width=0.98\columnwidth]{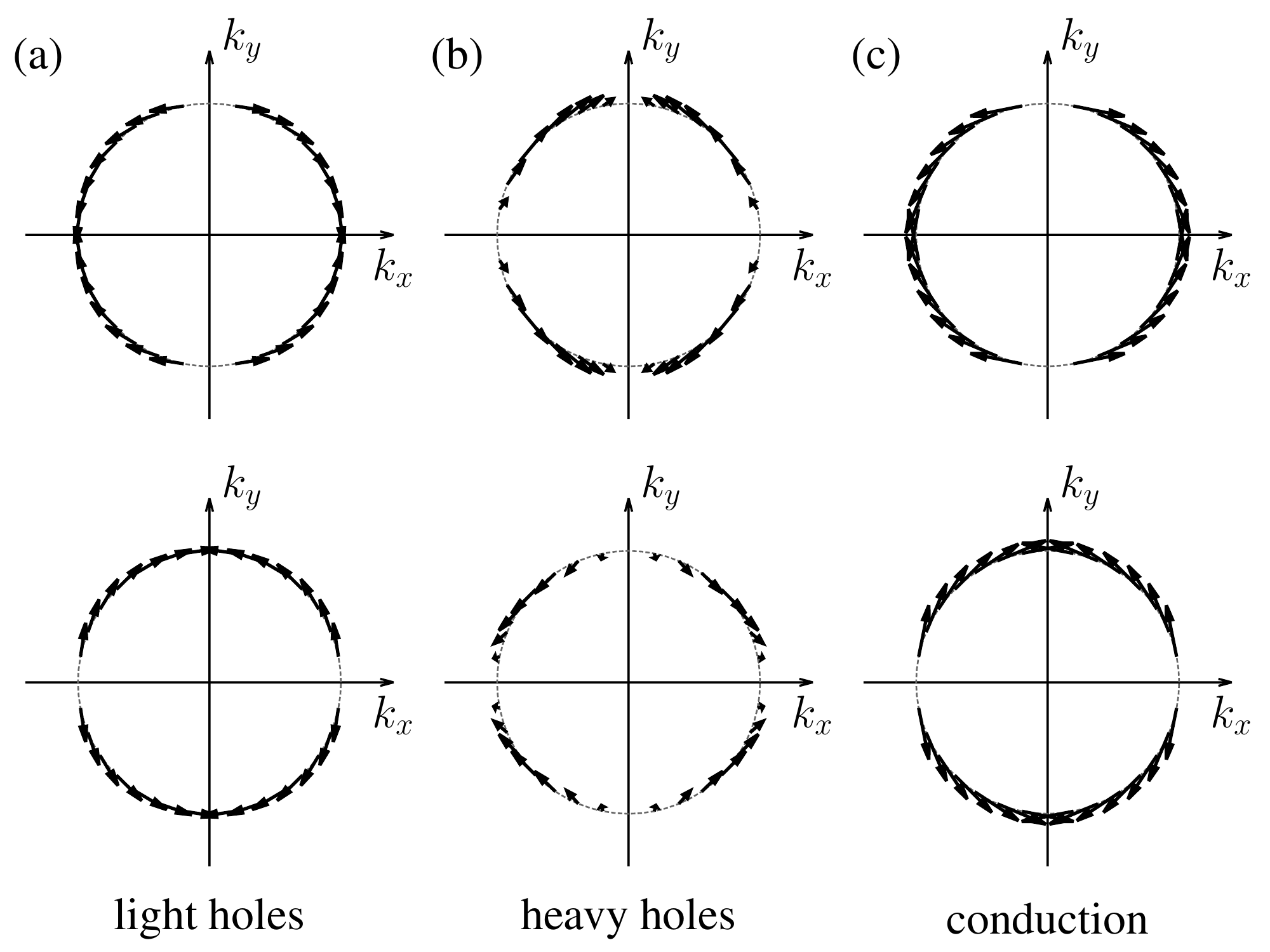}
	\caption{\label{Fig:App_spin_InAs}
		Calculated spin expectation values as in Fig.~\ref{Fig:App_spin_GaSb} but for zinc-blende InAs.
	}
\end{figure}

\begin{figure}[h!]
	\centering
	\includegraphics[width=0.98\columnwidth]{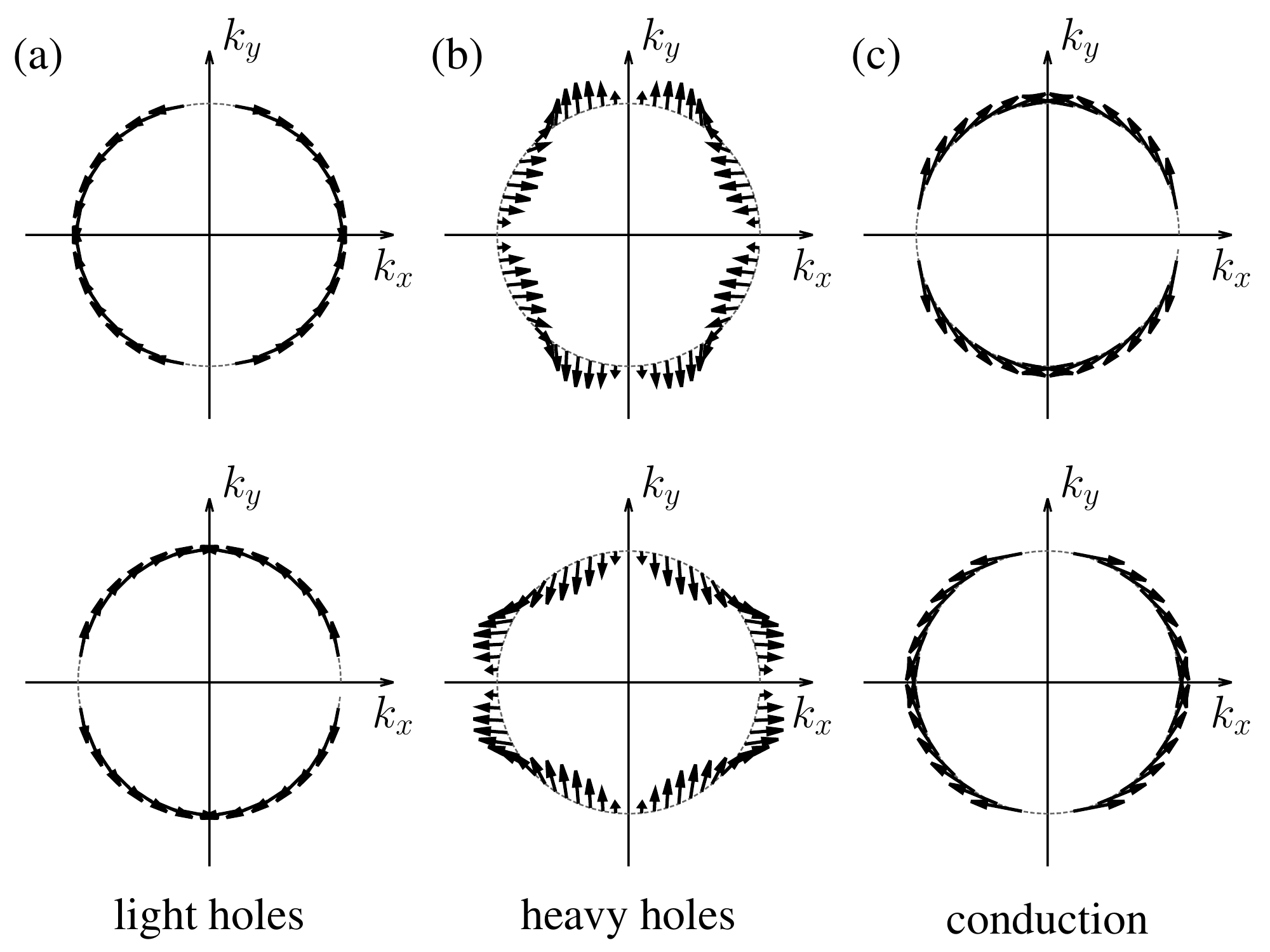}
	\caption{\label{Fig:App_spin_InSb}
		Calculated spin expectation values as in Fig.~\ref{Fig:App_spin_GaSb} but for zinc-blende InSb.
	}
\end{figure}

%-------------------------------------------------------------------
Here we present the calculated spin expectation values on a contour for momentum $k$ equal 1\% of Brillouin zone width, the $\Gamma - X$ length, for zinc-blende GaSb, Fig.~\ref{Fig:App_spin_GaSb}, InAs, Fig.~\ref{Fig:App_spin_InAs}, and InSb, Fig.~\ref{Fig:App_spin_InSb}. The spin for light holes and conduction bands are similar in all studied cases, compare also to GaAs in Fig.~\ref{Fig:spins_ZB}. Spins for the heavy holes in GaSb are zero for $k_x \approx k_y$, in addition to $k_x = 0$ and $k_y = 0$. 
We note that the amplitude of spins for the lower in energy band of heavy holes for InAs is about 0.65 smaller.

%------------------------------------------------------------------
%------------------------------------------------------------------

%------------------------------------------------------------------
%------------------------------------------------------------------

%------------------------------------------------------------------
%------------------------------------------------------------------

\bibliography{paper}

\begin{thebibliography}{89}%
\makeatletter
\providecommand \@ifxundefined [1]{%
 \@ifx{#1\undefined}
}%
\providecommand \@ifnum [1]{%
 \ifnum #1\expandafter \@firstoftwo
 \else \expandafter \@secondoftwo
 \fi
}%
\providecommand \@ifx [1]{%
 \ifx #1\expandafter \@firstoftwo
 \else \expandafter \@secondoftwo
 \fi
}%
\providecommand \natexlab [1]{#1}%
\providecommand \enquote  [1]{``#1''}%
\providecommand \bibnamefont  [1]{#1}%
\providecommand \bibfnamefont [1]{#1}%
\providecommand \citenamefont [1]{#1}%
\providecommand \href@noop [0]{\@secondoftwo}%
\providecommand \href [0]{\begingroup \@sanitize@url \@href}%
\providecommand \@href[1]{\@@startlink{#1}\@@href}%
\providecommand \@@href[1]{\endgroup#1\@@endlink}%
\providecommand \@sanitize@url [0]{\catcode `\\12\catcode `\$12\catcode
  `\&12\catcode `\#12\catcode `\^12\catcode `\_12\catcode `\%12\relax}%
\providecommand \@@startlink[1]{}%
\providecommand \@@endlink[0]{}%
\providecommand \url  [0]{\begingroup\@sanitize@url \@url }%
\providecommand \@url [1]{\endgroup\@href {#1}{\urlprefix }}%
\providecommand \urlprefix  [0]{URL }%
\providecommand \Eprint [0]{\href }%
\providecommand \doibase [0]{http://dx.doi.org/}%
\providecommand \selectlanguage [0]{\@gobble}%
\providecommand \bibinfo  [0]{\@secondoftwo}%
\providecommand \bibfield  [0]{\@secondoftwo}%
\providecommand \translation [1]{[#1]}%
\providecommand \BibitemOpen [0]{}%
\providecommand \bibitemStop [0]{}%
\providecommand \bibitemNoStop [0]{.\EOS\space}%
\providecommand \EOS [0]{\spacefactor3000\relax}%
\providecommand \BibitemShut  [1]{\csname bibitem#1\endcsname}%
\let\auto@bib@innerbib\@empty
%</preamble>
\bibitem [{\citenamefont {{\v{Z}uti{\'c}}}\ \emph {et~al.}(2004)\citenamefont
  {{\v{Z}uti{\'c}}}, \citenamefont {Fabian},\ and\ \citenamefont {{Das
  Sarma}}}]{Zutic2004:RMP}%
  \BibitemOpen
  \bibfield  {author} {\bibinfo {author} {\bibfnamefont {I.}~\bibnamefont
  {{\v{Z}uti{\'c}}}}, \bibinfo {author} {\bibfnamefont {J.}~\bibnamefont
  {Fabian}}, \ and\ \bibinfo {author} {\bibfnamefont {S.}~\bibnamefont {{Das
  Sarma}}},\ }\href {\doibase 10.1103/RevModPhys.76.323} {\bibfield  {journal}
  {\bibinfo  {journal} {Rev. Mod. Phys.}\ }\textbf {\bibinfo {volume} {76}},\
  \bibinfo {pages} {323} (\bibinfo {year} {2004})}\BibitemShut {NoStop}%
\bibitem [{\citenamefont {Fabian}\ \emph {et~al.}(2007)\citenamefont {Fabian},
  \citenamefont {Matos-Abiague}, \citenamefont {Ertler}, \citenamefont
  {Stano},\ and\ \citenamefont {\v{Z}uti\'c}}]{Fabian2007:APS}%
  \BibitemOpen
  \bibfield  {author} {\bibinfo {author} {\bibfnamefont {J.}~\bibnamefont
  {Fabian}}, \bibinfo {author} {\bibfnamefont {A.}~\bibnamefont
  {Matos-Abiague}}, \bibinfo {author} {\bibfnamefont {C.}~\bibnamefont
  {Ertler}}, \bibinfo {author} {\bibfnamefont {P.}~\bibnamefont {Stano}}, \
  and\ \bibinfo {author} {\bibfnamefont {I.}~\bibnamefont {\v{Z}uti\'c}},\
  }\href {\doibase http://www.physics.sk/aps/pub.php?y=2007&pub=aps-07-04}
  {\bibfield  {journal} {\bibinfo  {journal} {Acta Phys. Slovaca}\ }\textbf
  {\bibinfo {volume} {57}},\ \bibinfo {pages} {565} (\bibinfo {year}
  {2007})}\BibitemShut {NoStop}%
\bibitem [{\citenamefont {Dresselhaus}(1955)}]{Dresselhaus1955:PR}%
  \BibitemOpen
  \bibfield  {author} {\bibinfo {author} {\bibfnamefont {G.}~\bibnamefont
  {Dresselhaus}},\ }\href {\doibase 10.1103/PhysRev.100.580} {\bibfield
  {journal} {\bibinfo  {journal} {Phys. Rev.}\ }\textbf {\bibinfo {volume}
  {100}},\ \bibinfo {pages} {580} (\bibinfo {year} {1955})}\BibitemShut
  {NoStop}%
\bibitem [{\citenamefont {Fabian}\ and\ \citenamefont
  {Sarma}(1999)}]{Fabian1999:JVSTB}%
  \BibitemOpen
  \bibfield  {author} {\bibinfo {author} {\bibfnamefont {J.}~\bibnamefont
  {Fabian}}\ and\ \bibinfo {author} {\bibfnamefont {S.~D.}\ \bibnamefont
  {Sarma}},\ }\href {\doibase http://dx.doi.org/10.1116/1.590813} {\bibfield
  {journal} {\bibinfo  {journal} {Journal of Vacuum Science \& Technology B}\
  }\textbf {\bibinfo {volume} {17}},\ \bibinfo {pages} {1708} (\bibinfo {year}
  {1999})}\BibitemShut {NoStop}%
\bibitem [{\citenamefont {Meier}\ and\ \citenamefont
  {Zakharchenya}(1984)}]{Meier1984:book}%
  \BibitemOpen
  \bibfield  {author} {\bibinfo {author} {\bibfnamefont {F.}~\bibnamefont
  {Meier}}\ and\ \bibinfo {author} {\bibfnamefont {B.~P.}\ \bibnamefont
  {Zakharchenya}},\ }\href@noop {} {\emph {\bibinfo {title} {Optical
  orientation}}}\ (\bibinfo  {publisher} {Oxford, United Kingdom},\ \bibinfo
  {year} {1984})\BibitemShut {NoStop}%
\bibitem [{\citenamefont {Sinova}\ \emph {et~al.}(2015)\citenamefont {Sinova},
  \citenamefont {Valenzuela}, \citenamefont {Wunderlich}, \citenamefont
  {Back},\ and\ \citenamefont {Jungwirth}}]{Sinova2015:RMP}%
  \BibitemOpen
  \bibfield  {author} {\bibinfo {author} {\bibfnamefont {J.}~\bibnamefont
  {Sinova}}, \bibinfo {author} {\bibfnamefont {S.~O.}\ \bibnamefont
  {Valenzuela}}, \bibinfo {author} {\bibfnamefont {J.}~\bibnamefont
  {Wunderlich}}, \bibinfo {author} {\bibfnamefont {C.~H.}\ \bibnamefont
  {Back}}, \ and\ \bibinfo {author} {\bibfnamefont {T.}~\bibnamefont
  {Jungwirth}},\ }\href {\doibase 10.1103/RevModPhys.87.1213} {\bibfield
  {journal} {\bibinfo  {journal} {Rev. Mod. Phys.}\ }\textbf {\bibinfo {volume}
  {87}},\ \bibinfo {pages} {1213} (\bibinfo {year} {2015})}\BibitemShut
  {NoStop}%
\bibitem [{\citenamefont {Schliemann}(2016)}]{Schliemann2016}%
  \BibitemOpen
  \bibfield  {author} {\bibinfo {author} {\bibfnamefont {J.}~\bibnamefont
  {Schliemann}},\ }\href@noop {} {\bibfield  {journal} {\bibinfo  {journal}
  {arXiv:1604.02026}\ } (\bibinfo {year} {2016})}\BibitemShut {NoStop}%
\bibitem [{\citenamefont {Ganichev}(2008)}]{Ganichev2008:IJMPB}%
  \BibitemOpen
  \bibfield  {author} {\bibinfo {author} {\bibfnamefont {S.~D.}\ \bibnamefont
  {Ganichev}},\ }\href {\doibase dx.doi.org/10.1142/S0217979208046001}
  {\bibfield  {journal} {\bibinfo  {journal} {Int. J. Mod. Phys. B}\ }\textbf
  {\bibinfo {volume} {22}},\ \bibinfo {pages} {1} (\bibinfo {year}
  {2008})}\BibitemShut {NoStop}%
\bibitem [{\citenamefont {Lutchyn}\ \emph {et~al.}(2010)\citenamefont
  {Lutchyn}, \citenamefont {Sau},\ and\ \citenamefont
  {Das~Sarma}}]{Lutchyn2010:PRL}%
  \BibitemOpen
  \bibfield  {author} {\bibinfo {author} {\bibfnamefont {R.~M.}\ \bibnamefont
  {Lutchyn}}, \bibinfo {author} {\bibfnamefont {J.~D.}\ \bibnamefont {Sau}}, \
  and\ \bibinfo {author} {\bibfnamefont {S.}~\bibnamefont {Das~Sarma}},\ }\href
  {\doibase 10.1103/PhysRevLett.105.077001} {\bibfield  {journal} {\bibinfo
  {journal} {Phys. Rev. Lett.}\ }\textbf {\bibinfo {volume} {105}},\ \bibinfo
  {pages} {077001} (\bibinfo {year} {2010})}\BibitemShut {NoStop}%
\bibitem [{\citenamefont {Oreg}\ \emph {et~al.}(2010)\citenamefont {Oreg},
  \citenamefont {Refael},\ and\ \citenamefont {von Oppen}}]{Oreg2010:PRL}%
  \BibitemOpen
  \bibfield  {author} {\bibinfo {author} {\bibfnamefont {Y.}~\bibnamefont
  {Oreg}}, \bibinfo {author} {\bibfnamefont {G.}~\bibnamefont {Refael}}, \ and\
  \bibinfo {author} {\bibfnamefont {F.}~\bibnamefont {von Oppen}},\ }\href
  {\doibase 10.1103/PhysRevLett.105.177002} {\bibfield  {journal} {\bibinfo
  {journal} {Phys. Rev. Lett.}\ }\textbf {\bibinfo {volume} {105}},\ \bibinfo
  {pages} {177002} (\bibinfo {year} {2010})}\BibitemShut {NoStop}%
\bibitem [{\citenamefont {Mourik}\ \emph {et~al.}(2012)\citenamefont {Mourik},
  \citenamefont {Zuo}, \citenamefont {Frolov}, \citenamefont {Plissard},
  \citenamefont {Bakkers},\ and\ \citenamefont
  {Kouwenhoven}}]{Mourik2012:Science}%
  \BibitemOpen
  \bibfield  {author} {\bibinfo {author} {\bibfnamefont {V.}~\bibnamefont
  {Mourik}}, \bibinfo {author} {\bibfnamefont {K.}~\bibnamefont {Zuo}},
  \bibinfo {author} {\bibfnamefont {S.~M.}\ \bibnamefont {Frolov}}, \bibinfo
  {author} {\bibfnamefont {S.~R.}\ \bibnamefont {Plissard}}, \bibinfo {author}
  {\bibfnamefont {E.~P. A.~M.}\ \bibnamefont {Bakkers}}, \ and\ \bibinfo
  {author} {\bibfnamefont {L.~P.}\ \bibnamefont {Kouwenhoven}},\ }\href
  {\doibase 10.1126/science.1222360} {\bibfield  {journal} {\bibinfo  {journal}
  {Science}\ }\textbf {\bibinfo {volume} {336}},\ \bibinfo {pages} {1003}
  (\bibinfo {year} {2012})}\BibitemShut {NoStop}%
\bibitem [{\citenamefont {Deng}\ \emph {et~al.}(2012)\citenamefont {Deng},
  \citenamefont {Yu}, \citenamefont {Huang}, \citenamefont {Larsson},
  \citenamefont {Caroff},\ and\ \citenamefont {Xu}}]{Deng2012:NL}%
  \BibitemOpen
  \bibfield  {author} {\bibinfo {author} {\bibfnamefont {M.~T.}\ \bibnamefont
  {Deng}}, \bibinfo {author} {\bibfnamefont {C.~L.}\ \bibnamefont {Yu}},
  \bibinfo {author} {\bibfnamefont {G.~Y.}\ \bibnamefont {Huang}}, \bibinfo
  {author} {\bibfnamefont {M.}~\bibnamefont {Larsson}}, \bibinfo {author}
  {\bibfnamefont {P.}~\bibnamefont {Caroff}}, \ and\ \bibinfo {author}
  {\bibfnamefont {H.~Q.}\ \bibnamefont {Xu}},\ }\href {\doibase
  10.1021/nl303758w} {\bibfield  {journal} {\bibinfo  {journal} {Nano Letters}\
  }\textbf {\bibinfo {volume} {12}},\ \bibinfo {pages} {6414} (\bibinfo {year}
  {2012})}\BibitemShut {NoStop}%
\bibitem [{\citenamefont {Albrecht}\ \emph {et~al.}(2016)\citenamefont
  {Albrecht}, \citenamefont {Higginbotham}, \citenamefont {Madsen},
  \citenamefont {Kuemmeth}, \citenamefont {Jespersen}, \citenamefont
  {Nyg{\aa}rd}, \citenamefont {Krogstrup},\ and\ \citenamefont
  {Marcus}}]{Albrecht2016:Nature}%
  \BibitemOpen
  \bibfield  {author} {\bibinfo {author} {\bibfnamefont {S.~M.}\ \bibnamefont
  {Albrecht}}, \bibinfo {author} {\bibfnamefont {A.~P.}\ \bibnamefont
  {Higginbotham}}, \bibinfo {author} {\bibfnamefont {M.}~\bibnamefont
  {Madsen}}, \bibinfo {author} {\bibfnamefont {F.}~\bibnamefont {Kuemmeth}},
  \bibinfo {author} {\bibfnamefont {T.~S.}\ \bibnamefont {Jespersen}}, \bibinfo
  {author} {\bibfnamefont {J.}~\bibnamefont {Nyg{\aa}rd}}, \bibinfo {author}
  {\bibfnamefont {P.}~\bibnamefont {Krogstrup}}, \ and\ \bibinfo {author}
  {\bibfnamefont {C.~M.}\ \bibnamefont {Marcus}},\ }\href
  {http://dx.doi.org/10.1038/nature17162} {\bibfield  {journal} {\bibinfo
  {journal} {Nature}\ }\textbf {\bibinfo {volume} {531}},\ \bibinfo {pages}
  {206} (\bibinfo {year} {2016})},\ \bibinfo {note} {letter}\BibitemShut
  {NoStop}%
\bibitem [{\citenamefont {Das}\ \emph {et~al.}(2012)\citenamefont {Das},
  \citenamefont {Ronen}, \citenamefont {Most}, \citenamefont {Oreg},
  \citenamefont {Heiblum},\ and\ \citenamefont {Shtrikman}}]{Das2012:NP}%
  \BibitemOpen
  \bibfield  {author} {\bibinfo {author} {\bibfnamefont {A.}~\bibnamefont
  {Das}}, \bibinfo {author} {\bibfnamefont {Y.}~\bibnamefont {Ronen}}, \bibinfo
  {author} {\bibfnamefont {Y.}~\bibnamefont {Most}}, \bibinfo {author}
  {\bibfnamefont {Y.}~\bibnamefont {Oreg}}, \bibinfo {author} {\bibfnamefont
  {M.}~\bibnamefont {Heiblum}}, \ and\ \bibinfo {author} {\bibfnamefont
  {H.}~\bibnamefont {Shtrikman}},\ }\href {\doibase 10.1038/nphys2479}
  {\bibfield  {journal} {\bibinfo  {journal} {Nat Phys}\ }\textbf {\bibinfo
  {volume} {8}},\ \bibinfo {pages} {887} (\bibinfo {year} {2012})}\BibitemShut
  {NoStop}%
\bibitem [{\citenamefont {Cardona}\ \emph {et~al.}(1988)\citenamefont
  {Cardona}, \citenamefont {Christensen},\ and\ \citenamefont
  {Fasol}}]{Cardona1988:PRB}%
  \BibitemOpen
  \bibfield  {author} {\bibinfo {author} {\bibfnamefont {M.}~\bibnamefont
  {Cardona}}, \bibinfo {author} {\bibfnamefont {N.~E.}\ \bibnamefont
  {Christensen}}, \ and\ \bibinfo {author} {\bibfnamefont {G.}~\bibnamefont
  {Fasol}},\ }\href {\doibase 10.1103/PhysRevB.38.1806} {\bibfield  {journal}
  {\bibinfo  {journal} {Phys. Rev. B}\ }\textbf {\bibinfo {volume} {38}},\
  \bibinfo {pages} {1806} (\bibinfo {year} {1988})}\BibitemShut {NoStop}%
\bibitem [{\citenamefont {Winkler}(2003)}]{Winkler2003:book}%
  \BibitemOpen
  \bibfield  {author} {\bibinfo {author} {\bibfnamefont {R.}~\bibnamefont
  {Winkler}},\ }\href@noop {} {\emph {\bibinfo {title} {Spin-orbit coupling
  effects in two-dimensional electron and hole systems}}}\ (\bibinfo
  {publisher} {Springer, Berlin},\ \bibinfo {year} {2003})\BibitemShut
  {NoStop}%
\bibitem [{\citenamefont {Chantis}\ \emph
  {et~al.}(2006{\natexlab{a}})\citenamefont {Chantis}, \citenamefont {van
  Schilfgaarde},\ and\ \citenamefont {Kotani}}]{Chantis2006:PRL}%
  \BibitemOpen
  \bibfield  {author} {\bibinfo {author} {\bibfnamefont {A.~N.}\ \bibnamefont
  {Chantis}}, \bibinfo {author} {\bibfnamefont {M.}~\bibnamefont {van
  Schilfgaarde}}, \ and\ \bibinfo {author} {\bibfnamefont {T.}~\bibnamefont
  {Kotani}},\ }\href {\doibase 10.1103/PhysRevLett.96.086405} {\bibfield
  {journal} {\bibinfo  {journal} {Phys. Rev. Lett.}\ }\textbf {\bibinfo
  {volume} {96}},\ \bibinfo {pages} {086405} (\bibinfo {year}
  {2006}{\natexlab{a}})}\BibitemShut {NoStop}%
\bibitem [{\citenamefont {Caroff}\ \emph {et~al.}(2009)\citenamefont {Caroff},
  \citenamefont {Dick}, \citenamefont {Johansson}, \citenamefont {Messing},
  \citenamefont {Deppert},\ and\ \citenamefont {Samuelson}}]{Caroff2009:NN}%
  \BibitemOpen
  \bibfield  {author} {\bibinfo {author} {\bibfnamefont {P.}~\bibnamefont
  {Caroff}}, \bibinfo {author} {\bibfnamefont {K.~A.}\ \bibnamefont {Dick}},
  \bibinfo {author} {\bibfnamefont {J.}~\bibnamefont {Johansson}}, \bibinfo
  {author} {\bibfnamefont {M.~E.}\ \bibnamefont {Messing}}, \bibinfo {author}
  {\bibfnamefont {K.}~\bibnamefont {Deppert}}, \ and\ \bibinfo {author}
  {\bibfnamefont {L.}~\bibnamefont {Samuelson}},\ }\href {\doibase
  10.1038/nnano.2008.359} {\bibfield  {journal} {\bibinfo  {journal} {Nature
  Nanotechnology}\ }\textbf {\bibinfo {volume} {4}},\ \bibinfo {pages} {50}
  (\bibinfo {year} {2009})}\BibitemShut {NoStop}%
\bibitem [{\citenamefont {McMahon}\ and\ \citenamefont
  {Nelmes}(2005)}]{McMahon2005:PRL}%
  \BibitemOpen
  \bibfield  {author} {\bibinfo {author} {\bibfnamefont {M.~I.}\ \bibnamefont
  {McMahon}}\ and\ \bibinfo {author} {\bibfnamefont {R.~J.}\ \bibnamefont
  {Nelmes}},\ }\href {\doibase 10.1103/PhysRevLett.95.215505} {\bibfield
  {journal} {\bibinfo  {journal} {Phys. Rev. Lett.}\ }\textbf {\bibinfo
  {volume} {95}},\ \bibinfo {pages} {215505} (\bibinfo {year}
  {2005})}\BibitemShut {NoStop}%
\bibitem [{\citenamefont {Kriegner}\ \emph {et~al.}(2011)\citenamefont
  {Kriegner}, \citenamefont {Panse}, \citenamefont {Mandl}, \citenamefont
  {Dick}, \citenamefont {Keplinger}, \citenamefont {Persson}, \citenamefont
  {Caroff}, \citenamefont {Ercolani}, \citenamefont {Sorba}, \citenamefont
  {Bechstedt}, \citenamefont {Stangl},\ and\ \citenamefont
  {Bauer}}]{Kriegner2011:NL}%
  \BibitemOpen
  \bibfield  {author} {\bibinfo {author} {\bibfnamefont {D.}~\bibnamefont
  {Kriegner}}, \bibinfo {author} {\bibfnamefont {C.}~\bibnamefont {Panse}},
  \bibinfo {author} {\bibfnamefont {B.}~\bibnamefont {Mandl}}, \bibinfo
  {author} {\bibfnamefont {K.~A.}\ \bibnamefont {Dick}}, \bibinfo {author}
  {\bibfnamefont {M.}~\bibnamefont {Keplinger}}, \bibinfo {author}
  {\bibfnamefont {J.~M.}\ \bibnamefont {Persson}}, \bibinfo {author}
  {\bibfnamefont {P.}~\bibnamefont {Caroff}}, \bibinfo {author} {\bibfnamefont
  {D.}~\bibnamefont {Ercolani}}, \bibinfo {author} {\bibfnamefont
  {L.}~\bibnamefont {Sorba}}, \bibinfo {author} {\bibfnamefont
  {F.}~\bibnamefont {Bechstedt}}, \bibinfo {author} {\bibfnamefont
  {J.}~\bibnamefont {Stangl}}, \ and\ \bibinfo {author} {\bibfnamefont
  {G.}~\bibnamefont {Bauer}},\ }\href {\doibase 10.1021/nl1041512} {\bibfield
  {journal} {\bibinfo  {journal} {Nano Lett.}\ }\textbf {\bibinfo {volume}
  {11}},\ \bibinfo {pages} {1483} (\bibinfo {year} {2011})}\BibitemShut
  {NoStop}%
\bibitem [{\citenamefont {Zardo}\ \emph {et~al.}(2013)\citenamefont {Zardo},
  \citenamefont {Yazji}, \citenamefont {H{\"o}rmann}, \citenamefont
  {Hertenberger}, \citenamefont {Funk}, \citenamefont {Mangialardo},
  \citenamefont {Mork{\"o}tter}, \citenamefont {Koblm{\"u}ller}, \citenamefont
  {Postorino},\ and\ \citenamefont {Abstreiter}}]{Zardo2013:NL}%
  \BibitemOpen
  \bibfield  {author} {\bibinfo {author} {\bibfnamefont {I.}~\bibnamefont
  {Zardo}}, \bibinfo {author} {\bibfnamefont {S.}~\bibnamefont {Yazji}},
  \bibinfo {author} {\bibfnamefont {N.}~\bibnamefont {H{\"o}rmann}}, \bibinfo
  {author} {\bibfnamefont {S.}~\bibnamefont {Hertenberger}}, \bibinfo {author}
  {\bibfnamefont {S.}~\bibnamefont {Funk}}, \bibinfo {author} {\bibfnamefont
  {S.}~\bibnamefont {Mangialardo}}, \bibinfo {author} {\bibfnamefont
  {S.}~\bibnamefont {Mork{\"o}tter}}, \bibinfo {author} {\bibfnamefont
  {G.}~\bibnamefont {Koblm{\"u}ller}}, \bibinfo {author} {\bibfnamefont
  {P.}~\bibnamefont {Postorino}}, \ and\ \bibinfo {author} {\bibfnamefont
  {G.}~\bibnamefont {Abstreiter}},\ }\href {\doibase 10.1021/nl304528j}
  {\bibfield  {journal} {\bibinfo  {journal} {Nano Lett.}\ }\textbf {\bibinfo
  {volume} {13}},\ \bibinfo {pages} {3011} (\bibinfo {year}
  {2013})}\BibitemShut {NoStop}%
\bibitem [{\citenamefont {Dick}\ \emph {et~al.}(2010)\citenamefont {Dick},
  \citenamefont {Thelander}, \citenamefont {Samuelson},\ and\ \citenamefont
  {Caroff}}]{Dick2010:NL}%
  \BibitemOpen
  \bibfield  {author} {\bibinfo {author} {\bibfnamefont {K.~A.}\ \bibnamefont
  {Dick}}, \bibinfo {author} {\bibfnamefont {C.}~\bibnamefont {Thelander}},
  \bibinfo {author} {\bibfnamefont {L.}~\bibnamefont {Samuelson}}, \ and\
  \bibinfo {author} {\bibfnamefont {P.}~\bibnamefont {Caroff}},\ }\href
  {\doibase 10.1021/nl101632a} {\bibfield  {journal} {\bibinfo  {journal} {Nano
  Lett.}\ }\textbf {\bibinfo {volume} {10}},\ \bibinfo {pages} {3494} (\bibinfo
  {year} {2010})}\BibitemShut {NoStop}%
\bibitem [{\citenamefont {Joyce}\ \emph {et~al.}(2010)\citenamefont {Joyce},
  \citenamefont {Wong-Leung}, \citenamefont {Gao}, \citenamefont {Tan},\ and\
  \citenamefont {Jagadish}}]{Joyce2010:NL}%
  \BibitemOpen
  \bibfield  {author} {\bibinfo {author} {\bibfnamefont {H.~J.}\ \bibnamefont
  {Joyce}}, \bibinfo {author} {\bibfnamefont {J.}~\bibnamefont {Wong-Leung}},
  \bibinfo {author} {\bibfnamefont {Q.}~\bibnamefont {Gao}}, \bibinfo {author}
  {\bibfnamefont {H.~H.}\ \bibnamefont {Tan}}, \ and\ \bibinfo {author}
  {\bibfnamefont {C.}~\bibnamefont {Jagadish}},\ }\href {\doibase
  10.1021/nl903688v} {\bibfield  {journal} {\bibinfo  {journal} {Nano Lett.}\
  }\textbf {\bibinfo {volume} {10}},\ \bibinfo {pages} {908} (\bibinfo {year}
  {2010})}\BibitemShut {NoStop}%
\bibitem [{\citenamefont {Krogstrup}\ \emph {et~al.}(2010)\citenamefont
  {Krogstrup}, \citenamefont {Popovitz-Biro}, \citenamefont {Johnson},
  \citenamefont {Madsen}, \citenamefont {Nygård},\ and\ \citenamefont
  {Shtrikman}}]{Krogstrup2010:NL}%
  \BibitemOpen
  \bibfield  {author} {\bibinfo {author} {\bibfnamefont {P.}~\bibnamefont
  {Krogstrup}}, \bibinfo {author} {\bibfnamefont {R.}~\bibnamefont
  {Popovitz-Biro}}, \bibinfo {author} {\bibfnamefont {E.}~\bibnamefont
  {Johnson}}, \bibinfo {author} {\bibfnamefont {M.~H.}\ \bibnamefont {Madsen}},
  \bibinfo {author} {\bibfnamefont {J.}~\bibnamefont {Nygård}}, \ and\
  \bibinfo {author} {\bibfnamefont {H.}~\bibnamefont {Shtrikman}},\ }\href
  {\doibase 10.1021/nl102308k} {\bibfield  {journal} {\bibinfo  {journal} {Nano
  Lett.}\ }\textbf {\bibinfo {volume} {10}},\ \bibinfo {pages} {4475} (\bibinfo
  {year} {2010})}\BibitemShut {NoStop}%
\bibitem [{\citenamefont {Shtrikman}\ \emph
  {et~al.}(2009{\natexlab{a}})\citenamefont {Shtrikman}, \citenamefont
  {Popovitz-Biro}, \citenamefont {Kretinin},\ and\ \citenamefont
  {Heiblum}}]{Shtrikman2009:NL1}%
  \BibitemOpen
  \bibfield  {author} {\bibinfo {author} {\bibfnamefont {H.}~\bibnamefont
  {Shtrikman}}, \bibinfo {author} {\bibfnamefont {R.}~\bibnamefont
  {Popovitz-Biro}}, \bibinfo {author} {\bibfnamefont {A.}~\bibnamefont
  {Kretinin}}, \ and\ \bibinfo {author} {\bibfnamefont {M.}~\bibnamefont
  {Heiblum}},\ }\href {\doibase 10.1021/nl8027872} {\bibfield  {journal}
  {\bibinfo  {journal} {Nano Lett.}\ }\textbf {\bibinfo {volume} {9}},\
  \bibinfo {pages} {215} (\bibinfo {year} {2009}{\natexlab{a}})}\BibitemShut
  {NoStop}%
\bibitem [{\citenamefont {Shtrikman}\ \emph
  {et~al.}(2009{\natexlab{b}})\citenamefont {Shtrikman}, \citenamefont
  {Popovitz-Biro}, \citenamefont {Kretinin}, \citenamefont {Houben},
  \citenamefont {Heiblum}, \citenamefont {Buka\l{}a}, \citenamefont {Galicka},
  \citenamefont {Buczko},\ and\ \citenamefont {Kacman}}]{Shtrikman2009:NL2}%
  \BibitemOpen
  \bibfield  {author} {\bibinfo {author} {\bibfnamefont {H.}~\bibnamefont
  {Shtrikman}}, \bibinfo {author} {\bibfnamefont {R.}~\bibnamefont
  {Popovitz-Biro}}, \bibinfo {author} {\bibfnamefont {A.}~\bibnamefont
  {Kretinin}}, \bibinfo {author} {\bibfnamefont {L.}~\bibnamefont {Houben}},
  \bibinfo {author} {\bibfnamefont {M.}~\bibnamefont {Heiblum}}, \bibinfo
  {author} {\bibfnamefont {M.}~\bibnamefont {Buka\l{}a}}, \bibinfo {author}
  {\bibfnamefont {M.}~\bibnamefont {Galicka}}, \bibinfo {author} {\bibfnamefont
  {R.}~\bibnamefont {Buczko}}, \ and\ \bibinfo {author} {\bibfnamefont
  {P.}~\bibnamefont {Kacman}},\ }\href {\doibase 10.1021/nl803524s} {\bibfield
  {journal} {\bibinfo  {journal} {Nano Lett.}\ }\textbf {\bibinfo {volume}
  {9}},\ \bibinfo {pages} {1506} (\bibinfo {year}
  {2009}{\natexlab{b}})}\BibitemShut {NoStop}%
\bibitem [{\citenamefont {Krogstrup}\ \emph {et~al.}(2015)\citenamefont
  {Krogstrup}, \citenamefont {Albrecht}, \citenamefont {Jespersen},
  \citenamefont {Marcus}, \citenamefont {Nygård}, \citenamefont {Johnson},
  \citenamefont {Madsen}, \citenamefont {Chang}, \citenamefont {Ziino},
  \citenamefont {Ziino}, \citenamefont {Chang}, \citenamefont {Albrecht},
  \citenamefont {Madsen}, \citenamefont {Johnson}, \citenamefont {Nygård},
  \citenamefont {Marcus},\ and\ \citenamefont {Jespersen}}]{Krogstrup2015:NM}%
  \BibitemOpen
  \bibfield  {author} {\bibinfo {author} {\bibfnamefont {P.}~\bibnamefont
  {Krogstrup}}, \bibinfo {author} {\bibfnamefont {S.~M.}\ \bibnamefont
  {Albrecht}}, \bibinfo {author} {\bibfnamefont {T.~S.}\ \bibnamefont
  {Jespersen}}, \bibinfo {author} {\bibfnamefont {C.~M.}\ \bibnamefont
  {Marcus}}, \bibinfo {author} {\bibfnamefont {J.}~\bibnamefont {Nygård}},
  \bibinfo {author} {\bibfnamefont {E.}~\bibnamefont {Johnson}}, \bibinfo
  {author} {\bibfnamefont {M.~H.}\ \bibnamefont {Madsen}}, \bibinfo {author}
  {\bibfnamefont {W.}~\bibnamefont {Chang}}, \bibinfo {author} {\bibfnamefont
  {N.~L.~B.}\ \bibnamefont {Ziino}}, \bibinfo {author} {\bibfnamefont
  {N.~L.~B.}\ \bibnamefont {Ziino}}, \bibinfo {author} {\bibfnamefont
  {W.}~\bibnamefont {Chang}}, \bibinfo {author} {\bibfnamefont {S.~M.}\
  \bibnamefont {Albrecht}}, \bibinfo {author} {\bibfnamefont {M.~H.}\
  \bibnamefont {Madsen}}, \bibinfo {author} {\bibfnamefont {E.}~\bibnamefont
  {Johnson}}, \bibinfo {author} {\bibfnamefont {J.}~\bibnamefont {Nygård}},
  \bibinfo {author} {\bibfnamefont {C.~M.}\ \bibnamefont {Marcus}}, \ and\
  \bibinfo {author} {\bibfnamefont {T.~S.}\ \bibnamefont {Jespersen}},\ }\href
  {\doibase 10.1038/nmat4176} {\bibfield  {journal} {\bibinfo  {journal} {Nat
  Mater}\ }\textbf {\bibinfo {volume} {advance online publication}} (\bibinfo
  {year} {2015}),\ 10.1038/nmat4176}\BibitemShut {NoStop}%
\bibitem [{\citenamefont {Wilhelm}\ \emph {et~al.}(2012)\citenamefont
  {Wilhelm}, \citenamefont {Larrue}, \citenamefont {Dai}, \citenamefont
  {Migas}, \citenamefont {Soci}, \citenamefont {Larrue}, \citenamefont {Dai},
  \citenamefont {Wilhelm},\ and\ \citenamefont {Soci}}]{Wilhelm2012:NS}%
  \BibitemOpen
  \bibfield  {author} {\bibinfo {author} {\bibfnamefont {C.}~\bibnamefont
  {Wilhelm}}, \bibinfo {author} {\bibfnamefont {A.}~\bibnamefont {Larrue}},
  \bibinfo {author} {\bibfnamefont {X.}~\bibnamefont {Dai}}, \bibinfo {author}
  {\bibfnamefont {D.}~\bibnamefont {Migas}}, \bibinfo {author} {\bibfnamefont
  {C.}~\bibnamefont {Soci}}, \bibinfo {author} {\bibfnamefont {A.}~\bibnamefont
  {Larrue}}, \bibinfo {author} {\bibfnamefont {X.}~\bibnamefont {Dai}},
  \bibinfo {author} {\bibfnamefont {C.}~\bibnamefont {Wilhelm}}, \ and\
  \bibinfo {author} {\bibfnamefont {C.}~\bibnamefont {Soci}},\ }\href {\doibase
  10.1039/C2NR00045H} {\bibfield  {journal} {\bibinfo  {journal} {Nanoscale}\
  }\textbf {\bibinfo {volume} {4}},\ \bibinfo {pages} {1446} (\bibinfo {year}
  {2012})}\BibitemShut {NoStop}%
\bibitem [{\citenamefont {Furthmeier}\ \emph {et~al.}(2016)\citenamefont
  {Furthmeier}, \citenamefont {Dirnberger}, \citenamefont {Gmitra},
  \citenamefont {Bayer}, \citenamefont {Hubmann}, \citenamefont {Sch{\"u}ller},
  \citenamefont {Reiger}, \citenamefont {Fabian}, \citenamefont {Korn},\ and\
  \citenamefont {Bougeard}}]{Furtmeier2016}%
  \BibitemOpen
  \bibfield  {author} {\bibinfo {author} {\bibfnamefont {S.}~\bibnamefont
  {Furthmeier}}, \bibinfo {author} {\bibfnamefont {F.}~\bibnamefont
  {Dirnberger}}, \bibinfo {author} {\bibfnamefont {M.}~\bibnamefont {Gmitra}},
  \bibinfo {author} {\bibfnamefont {A.}~\bibnamefont {Bayer}}, \bibinfo
  {author} {\bibfnamefont {J.}~\bibnamefont {Hubmann}}, \bibinfo {author}
  {\bibfnamefont {C.}~\bibnamefont {Sch{\"u}ller}}, \bibinfo {author}
  {\bibfnamefont {E.}~\bibnamefont {Reiger}}, \bibinfo {author} {\bibfnamefont
  {J.}~\bibnamefont {Fabian}}, \bibinfo {author} {\bibfnamefont
  {T.}~\bibnamefont {Korn}}, \ and\ \bibinfo {author} {\bibfnamefont
  {D.}~\bibnamefont {Bougeard}},\ }\href@noop {} {\bibfield  {journal}
  {\bibinfo  {journal} {Nat. Commun.}\ } (\bibinfo {year} {2016})}\BibitemShut
  {NoStop}%
\bibitem [{\citenamefont {Chelikowsky}\ and\ \citenamefont
  {Cohen}(1976)}]{Chelikowsky1976:PRB}%
  \BibitemOpen
  \bibfield  {author} {\bibinfo {author} {\bibfnamefont {J.~R.}\ \bibnamefont
  {Chelikowsky}}\ and\ \bibinfo {author} {\bibfnamefont {M.~L.}\ \bibnamefont
  {Cohen}},\ }\href {\doibase 10.1103/PhysRevB.14.556} {\bibfield  {journal}
  {\bibinfo  {journal} {Phys. Rev. B}\ }\textbf {\bibinfo {volume} {14}},\
  \bibinfo {pages} {556} (\bibinfo {year} {1976})}\BibitemShut {NoStop}%
\bibitem [{\citenamefont {Hohenberg}\ and\ \citenamefont
  {Kohn}(1964)}]{Hohenberg1964:PR}%
  \BibitemOpen
  \bibfield  {author} {\bibinfo {author} {\bibfnamefont {J.}~\bibnamefont
  {Hohenberg}}\ and\ \bibinfo {author} {\bibfnamefont {W.}~\bibnamefont
  {Kohn}},\ }\href {\doibase 10.1103/PhysRev.136.B864} {\bibfield  {journal}
  {\bibinfo  {journal} {Phys. Rev.}\ }\textbf {\bibinfo {volume} {136}},\
  \bibinfo {pages} {B864} (\bibinfo {year} {1964})}\BibitemShut {NoStop}%
\bibitem [{\citenamefont {Murayama}\ and\ \citenamefont
  {Nakayama}(1994)}]{Murayama1994:PRB}%
  \BibitemOpen
  \bibfield  {author} {\bibinfo {author} {\bibfnamefont {M.}~\bibnamefont
  {Murayama}}\ and\ \bibinfo {author} {\bibfnamefont {T.}~\bibnamefont
  {Nakayama}},\ }\href {\doibase 10.1103/PhysRevB.49.4710} {\bibfield
  {journal} {\bibinfo  {journal} {Phys. Rev. B}\ }\textbf {\bibinfo {volume}
  {49}},\ \bibinfo {pages} {4710} (\bibinfo {year} {1994})}\BibitemShut
  {NoStop}%
\bibitem [{\citenamefont {Perdew}\ and\ \citenamefont
  {Zunger}(1981)}]{Perdew1981:PRB}%
  \BibitemOpen
  \bibfield  {author} {\bibinfo {author} {\bibfnamefont {J.~P.}\ \bibnamefont
  {Perdew}}\ and\ \bibinfo {author} {\bibfnamefont {A.}~\bibnamefont
  {Zunger}},\ }\href {\doibase 10.1103/PhysRevB.23.5048} {\bibfield  {journal}
  {\bibinfo  {journal} {Phys. Rev. B}\ }\textbf {\bibinfo {volume} {23}},\
  \bibinfo {pages} {5048} (\bibinfo {year} {1981})}\BibitemShut {NoStop}%
\bibitem [{\citenamefont {Chantis}\ \emph
  {et~al.}(2006{\natexlab{b}})\citenamefont {Chantis}, \citenamefont {van
  Schilfgaarde},\ and\ \citenamefont {Kotani}}]{Chantis2006:PRL-erratum}%
  \BibitemOpen
  \bibfield  {author} {\bibinfo {author} {\bibfnamefont {A.~N.}\ \bibnamefont
  {Chantis}}, \bibinfo {author} {\bibfnamefont {M.}~\bibnamefont {van
  Schilfgaarde}}, \ and\ \bibinfo {author} {\bibfnamefont {T.}~\bibnamefont
  {Kotani}},\ }\href {\doibase 10.1103/PhysRevLett.97.039903.} {\bibfield
  {journal} {\bibinfo  {journal} {Phys. Rev. Lett.}\ }\textbf {\bibinfo
  {volume} {97}},\ \bibinfo {pages} {039903(E)} (\bibinfo {year}
  {2006}{\natexlab{b}})}\BibitemShut {NoStop}%
\bibitem [{\citenamefont {Becke}\ and\ \citenamefont
  {Johnson}(2006)}]{Becke2006:JCHP}%
  \BibitemOpen
  \bibfield  {author} {\bibinfo {author} {\bibfnamefont {A.~D.}\ \bibnamefont
  {Becke}}\ and\ \bibinfo {author} {\bibfnamefont {E.~R.}\ \bibnamefont
  {Johnson}},\ }\href {\doibase 10.1063/1.2213970} {\bibfield  {journal}
  {\bibinfo  {journal} {The Journal of Chemical Physics}\ }\textbf {\bibinfo
  {volume} {124}},\ \bibinfo {pages} {221101} (\bibinfo {year}
  {2006})}\BibitemShut {NoStop}%
\bibitem [{\citenamefont {Tran}\ and\ \citenamefont
  {Blaha}(2009)}]{Tran2009:PRL}%
  \BibitemOpen
  \bibfield  {author} {\bibinfo {author} {\bibfnamefont {F.}~\bibnamefont
  {Tran}}\ and\ \bibinfo {author} {\bibfnamefont {P.}~\bibnamefont {Blaha}},\
  }\href {\doibase 10.1103/PhysRevLett.102.226401} {\bibfield  {journal}
  {\bibinfo  {journal} {Phys. Rev. Lett.}\ }\textbf {\bibinfo {volume} {102}},\
  \bibinfo {pages} {226401} (\bibinfo {year} {2009})}\BibitemShut {NoStop}%
\bibitem [{\citenamefont {Soluyanov}\ \emph {et~al.}(2016)\citenamefont
  {Soluyanov}, \citenamefont {Gresch}, \citenamefont {Troyer}, \citenamefont
  {Lutchyn}, \citenamefont {Bauer},\ and\ \citenamefont
  {Nayak}}]{Soluyanov2016:PRB}%
  \BibitemOpen
  \bibfield  {author} {\bibinfo {author} {\bibfnamefont {A.~A.}\ \bibnamefont
  {Soluyanov}}, \bibinfo {author} {\bibfnamefont {D.}~\bibnamefont {Gresch}},
  \bibinfo {author} {\bibfnamefont {M.}~\bibnamefont {Troyer}}, \bibinfo
  {author} {\bibfnamefont {R.~M.}\ \bibnamefont {Lutchyn}}, \bibinfo {author}
  {\bibfnamefont {B.}~\bibnamefont {Bauer}}, \ and\ \bibinfo {author}
  {\bibfnamefont {C.}~\bibnamefont {Nayak}},\ }\href {\doibase
  10.1103/PhysRevB.93.115317} {\bibfield  {journal} {\bibinfo  {journal} {Phys.
  Rev. B}\ }\textbf {\bibinfo {volume} {93}},\ \bibinfo {pages} {115317}
  (\bibinfo {year} {2016})}\BibitemShut {NoStop}%
\bibitem [{\citenamefont {Junior}\ \emph {et~al.}(2016)\citenamefont {Junior},
  \citenamefont {Campos}, \citenamefont {Bastos}, \citenamefont {Gmitra},
  \citenamefont {Fabian},\ and\ \citenamefont {Sipahi}}]{Paulo2016:arXiv}%
  \BibitemOpen
  \bibfield  {author} {\bibinfo {author} {\bibfnamefont {P.~E.~F.}\
  \bibnamefont {Junior}}, \bibinfo {author} {\bibfnamefont {T.}~\bibnamefont
  {Campos}}, \bibinfo {author} {\bibfnamefont {C.~M.~O.}\ \bibnamefont
  {Bastos}}, \bibinfo {author} {\bibfnamefont {M.}~\bibnamefont {Gmitra}},
  \bibinfo {author} {\bibfnamefont {J.}~\bibnamefont {Fabian}}, \ and\ \bibinfo
  {author} {\bibfnamefont {G.~M.}\ \bibnamefont {Sipahi}},\ }\href@noop {}
  {\bibfield  {journal} {\bibinfo  {journal} {arXiv:1604.06014}\ } (\bibinfo
  {year} {2016})}\BibitemShut {NoStop}%
\bibitem [{\citenamefont {K\"ackell}\ \emph {et~al.}(1994)\citenamefont
  {K\"ackell}, \citenamefont {Wenzien},\ and\ \citenamefont
  {Bechstedt}}]{Kackell1994:PRB}%
  \BibitemOpen
  \bibfield  {author} {\bibinfo {author} {\bibfnamefont {P.}~\bibnamefont
  {K\"ackell}}, \bibinfo {author} {\bibfnamefont {B.}~\bibnamefont {Wenzien}},
  \ and\ \bibinfo {author} {\bibfnamefont {F.}~\bibnamefont {Bechstedt}},\
  }\href {\doibase 10.1103/PhysRevB.50.17037} {\bibfield  {journal} {\bibinfo
  {journal} {Phys. Rev. B}\ }\textbf {\bibinfo {volume} {50}},\ \bibinfo
  {pages} {17037} (\bibinfo {year} {1994})}\BibitemShut {NoStop}%
\bibitem [{\citenamefont {Belabbes}\ \emph {et~al.}(2012)\citenamefont
  {Belabbes}, \citenamefont {Panse}, \citenamefont {Furthm{\"u}ller},\ and\
  \citenamefont {Bechstedt}}]{Belabbes2012:PRB}%
  \BibitemOpen
  \bibfield  {author} {\bibinfo {author} {\bibfnamefont {A.}~\bibnamefont
  {Belabbes}}, \bibinfo {author} {\bibfnamefont {C.}~\bibnamefont {Panse}},
  \bibinfo {author} {\bibfnamefont {J.}~\bibnamefont {Furthm{\"u}ller}}, \ and\
  \bibinfo {author} {\bibfnamefont {F.}~\bibnamefont {Bechstedt}},\ }\href
  {\doibase 10.1103/PhysRevB.86.075208} {\bibfield  {journal} {\bibinfo
  {journal} {Phys. Rev. B}\ }\textbf {\bibinfo {volume} {86}},\ \bibinfo
  {pages} {075208} (\bibinfo {year} {2012})}\BibitemShut {NoStop}%
\bibitem [{\citenamefont {Chantis}\ \emph {et~al.}(2010)\citenamefont
  {Chantis}, \citenamefont {Christensen}, \citenamefont {Svane},\ and\
  \citenamefont {Cardona}}]{Chantis2010:PRB}%
  \BibitemOpen
  \bibfield  {author} {\bibinfo {author} {\bibfnamefont {A.~N.}\ \bibnamefont
  {Chantis}}, \bibinfo {author} {\bibfnamefont {N.~E.}\ \bibnamefont
  {Christensen}}, \bibinfo {author} {\bibfnamefont {A.}~\bibnamefont {Svane}},
  \ and\ \bibinfo {author} {\bibfnamefont {M.}~\bibnamefont {Cardona}},\ }\href
  {\doibase 10.1103/PhysRevB.81.205205} {\bibfield  {journal} {\bibinfo
  {journal} {Phys. Rev. B}\ }\textbf {\bibinfo {volume} {81}},\ \bibinfo
  {pages} {205205} (\bibinfo {year} {2010})}\BibitemShut {NoStop}%
\bibitem [{\citenamefont {Zanolli}\ \emph
  {et~al.}(2007{\natexlab{a}})\citenamefont {Zanolli}, \citenamefont {Fuchs},
  \citenamefont {Furthm{\"u}ller}, \citenamefont {von Barth},\ and\
  \citenamefont {Bechstedt}}]{Zanolli2007:PRB}%
  \BibitemOpen
  \bibfield  {author} {\bibinfo {author} {\bibfnamefont {Z.}~\bibnamefont
  {Zanolli}}, \bibinfo {author} {\bibfnamefont {F.}~\bibnamefont {Fuchs}},
  \bibinfo {author} {\bibfnamefont {J.}~\bibnamefont {Furthm{\"u}ller}},
  \bibinfo {author} {\bibfnamefont {U.}~\bibnamefont {von Barth}}, \ and\
  \bibinfo {author} {\bibfnamefont {F.}~\bibnamefont {Bechstedt}},\ }\href
  {\doibase 10.1103/PhysRevB.75.245121} {\bibfield  {journal} {\bibinfo
  {journal} {Phys. Rev. B}\ }\textbf {\bibinfo {volume} {75}},\ \bibinfo
  {pages} {245121} (\bibinfo {year} {2007}{\natexlab{a}})}\BibitemShut
  {NoStop}%
\bibitem [{\citenamefont {Cheiwchanchamnangij}\ and\ \citenamefont
  {Lambrecht}(2011)}]{Cheiwchanchamnangij2011:PRB}%
  \BibitemOpen
  \bibfield  {author} {\bibinfo {author} {\bibfnamefont {T.}~\bibnamefont
  {Cheiwchanchamnangij}}\ and\ \bibinfo {author} {\bibfnamefont {W.~R.~L.}\
  \bibnamefont {Lambrecht}},\ }\href {\doibase 10.1103/PhysRevB.84.035203}
  {\bibfield  {journal} {\bibinfo  {journal} {Phys. Rev. B}\ }\textbf {\bibinfo
  {volume} {84}},\ \bibinfo {pages} {035203} (\bibinfo {year}
  {2011})}\BibitemShut {NoStop}%
\bibitem [{\citenamefont {Panse}\ \emph {et~al.}(2011)\citenamefont {Panse},
  \citenamefont {Kriegner},\ and\ \citenamefont {Bechstedt}}]{Panse2011:PRB}%
  \BibitemOpen
  \bibfield  {author} {\bibinfo {author} {\bibfnamefont {C.}~\bibnamefont
  {Panse}}, \bibinfo {author} {\bibfnamefont {D.}~\bibnamefont {Kriegner}}, \
  and\ \bibinfo {author} {\bibfnamefont {F.}~\bibnamefont {Bechstedt}},\ }\href
  {\doibase 10.1103/PhysRevB.84.075217} {\bibfield  {journal} {\bibinfo
  {journal} {Phys. Rev. B}\ }\textbf {\bibinfo {volume} {84}},\ \bibinfo
  {pages} {075217} (\bibinfo {year} {2011})}\BibitemShut {NoStop}%
\bibitem [{\citenamefont {Ferreira}\ \emph {et~al.}(2008)\citenamefont
  {Ferreira}, \citenamefont {Marques},\ and\ \citenamefont
  {Teles}}]{Ferreira2008:PRB}%
  \BibitemOpen
  \bibfield  {author} {\bibinfo {author} {\bibfnamefont {L.~G.}\ \bibnamefont
  {Ferreira}}, \bibinfo {author} {\bibfnamefont {M.}~\bibnamefont {Marques}}, \
  and\ \bibinfo {author} {\bibfnamefont {L.~K.}\ \bibnamefont {Teles}},\ }\href
  {\doibase 10.1103/PhysRevB.78.125116} {\bibfield  {journal} {\bibinfo
  {journal} {Phys. Rev. B}\ }\textbf {\bibinfo {volume} {78}},\ \bibinfo
  {pages} {125116} (\bibinfo {year} {2008})}\BibitemShut {NoStop}%
\bibitem [{\citenamefont {Betzinger}\ \emph {et~al.}(2010)\citenamefont
  {Betzinger}, \citenamefont {Friedrich},\ and\ \citenamefont
  {Bl\"ugel}}]{Betzinger2010:PRB}%
  \BibitemOpen
  \bibfield  {author} {\bibinfo {author} {\bibfnamefont {M.}~\bibnamefont
  {Betzinger}}, \bibinfo {author} {\bibfnamefont {C.}~\bibnamefont
  {Friedrich}}, \ and\ \bibinfo {author} {\bibfnamefont {S.}~\bibnamefont
  {Bl\"ugel}},\ }\href {\doibase 10.1103/PhysRevB.81.195117} {\bibfield
  {journal} {\bibinfo  {journal} {Phys. Rev. B}\ }\textbf {\bibinfo {volume}
  {81}},\ \bibinfo {pages} {195117} (\bibinfo {year} {2010})}\BibitemShut
  {NoStop}%
\bibitem [{\citenamefont {Tran}\ and\ \citenamefont
  {Blaha}(2011)}]{Tran2011:PRB}%
  \BibitemOpen
  \bibfield  {author} {\bibinfo {author} {\bibfnamefont {F.}~\bibnamefont
  {Tran}}\ and\ \bibinfo {author} {\bibfnamefont {P.}~\bibnamefont {Blaha}},\
  }\href {\doibase 10.1103/PhysRevB.83.235118} {\bibfield  {journal} {\bibinfo
  {journal} {Phys. Rev. B}\ }\textbf {\bibinfo {volume} {83}},\ \bibinfo
  {pages} {235118} (\bibinfo {year} {2011})}\BibitemShut {NoStop}%
\bibitem [{\citenamefont {Friedrich}\ \emph {et~al.}(2012)\citenamefont
  {Friedrich}, \citenamefont {Betzinger}, \citenamefont {Schlipf},
  \citenamefont {Bl{\"u}gel},\ and\ \citenamefont
  {Schindlmayr}}]{Friedrich2012:JPCM}%
  \BibitemOpen
  \bibfield  {author} {\bibinfo {author} {\bibfnamefont {C.}~\bibnamefont
  {Friedrich}}, \bibinfo {author} {\bibfnamefont {M.}~\bibnamefont
  {Betzinger}}, \bibinfo {author} {\bibfnamefont {M.}~\bibnamefont {Schlipf}},
  \bibinfo {author} {\bibfnamefont {S.}~\bibnamefont {Bl{\"u}gel}}, \ and\
  \bibinfo {author} {\bibfnamefont {A.}~\bibnamefont {Schindlmayr}},\ }\href
  {http://stacks.iop.org/0953-8984/24/i=29/a=293201} {\bibfield  {journal}
  {\bibinfo  {journal} {Journal of Physics: Condensed Matter}\ }\textbf
  {\bibinfo {volume} {24}},\ \bibinfo {pages} {293201} (\bibinfo {year}
  {2012})}\BibitemShut {NoStop}%
\bibitem [{\citenamefont {Heiss}\ \emph {et~al.}(2011)\citenamefont {Heiss},
  \citenamefont {Conesa-Boj}, \citenamefont {Ren}, \citenamefont {Tseng},
  \citenamefont {Gali}, \citenamefont {Rudolph}, \citenamefont {Uccelli},
  \citenamefont {Peir\'o}, \citenamefont {Morante}, \citenamefont {Schuh},
  \citenamefont {Reiger}, \citenamefont {Kaxiras}, \citenamefont {Arbiol},\
  and\ \citenamefont {Fontcuberta~i Morral}}]{Heiss2011:PRB}%
  \BibitemOpen
  \bibfield  {author} {\bibinfo {author} {\bibfnamefont {M.}~\bibnamefont
  {Heiss}}, \bibinfo {author} {\bibfnamefont {S.}~\bibnamefont {Conesa-Boj}},
  \bibinfo {author} {\bibfnamefont {J.}~\bibnamefont {Ren}}, \bibinfo {author}
  {\bibfnamefont {H.-H.}\ \bibnamefont {Tseng}}, \bibinfo {author}
  {\bibfnamefont {A.}~\bibnamefont {Gali}}, \bibinfo {author} {\bibfnamefont
  {A.}~\bibnamefont {Rudolph}}, \bibinfo {author} {\bibfnamefont
  {E.}~\bibnamefont {Uccelli}}, \bibinfo {author} {\bibfnamefont
  {F.}~\bibnamefont {Peir\'o}}, \bibinfo {author} {\bibfnamefont {J.~R.}\
  \bibnamefont {Morante}}, \bibinfo {author} {\bibfnamefont {D.}~\bibnamefont
  {Schuh}}, \bibinfo {author} {\bibfnamefont {E.}~\bibnamefont {Reiger}},
  \bibinfo {author} {\bibfnamefont {E.}~\bibnamefont {Kaxiras}}, \bibinfo
  {author} {\bibfnamefont {J.}~\bibnamefont {Arbiol}}, \ and\ \bibinfo {author}
  {\bibfnamefont {A.}~\bibnamefont {Fontcuberta~i Morral}},\ }\href {\doibase
  10.1103/PhysRevB.83.045303} {\bibfield  {journal} {\bibinfo  {journal} {Phys.
  Rev. B}\ }\textbf {\bibinfo {volume} {83}},\ \bibinfo {pages} {045303}
  (\bibinfo {year} {2011})}\BibitemShut {NoStop}%
\bibitem [{\citenamefont {Kim}\ \emph {et~al.}(2010)\citenamefont {Kim},
  \citenamefont {Marsman}, \citenamefont {Kresse}, \citenamefont {Tran},\ and\
  \citenamefont {Blaha}}]{Kim2010:PRB}%
  \BibitemOpen
  \bibfield  {author} {\bibinfo {author} {\bibfnamefont {Y.-S.}\ \bibnamefont
  {Kim}}, \bibinfo {author} {\bibfnamefont {M.}~\bibnamefont {Marsman}},
  \bibinfo {author} {\bibfnamefont {G.}~\bibnamefont {Kresse}}, \bibinfo
  {author} {\bibfnamefont {F.}~\bibnamefont {Tran}}, \ and\ \bibinfo {author}
  {\bibfnamefont {P.}~\bibnamefont {Blaha}},\ }\href {\doibase
  10.1103/PhysRevB.82.205212} {\bibfield  {journal} {\bibinfo  {journal} {Phys.
  Rev. B}\ }\textbf {\bibinfo {volume} {82}},\ \bibinfo {pages} {205212}
  (\bibinfo {year} {2010})}\BibitemShut {NoStop}%
\bibitem [{\citenamefont {Koller}\ \emph {et~al.}(2011)\citenamefont {Koller},
  \citenamefont {Tran},\ and\ \citenamefont {Blaha}}]{Koller2011:PRB}%
  \BibitemOpen
  \bibfield  {author} {\bibinfo {author} {\bibfnamefont {D.}~\bibnamefont
  {Koller}}, \bibinfo {author} {\bibfnamefont {F.}~\bibnamefont {Tran}}, \ and\
  \bibinfo {author} {\bibfnamefont {P.}~\bibnamefont {Blaha}},\ }\href
  {\doibase 10.1103/PhysRevB.83.195134} {\bibfield  {journal} {\bibinfo
  {journal} {Phys. Rev. B}\ }\textbf {\bibinfo {volume} {83}},\ \bibinfo
  {pages} {195134} (\bibinfo {year} {2011})}\BibitemShut {NoStop}%
\bibitem [{\citenamefont {Koller}\ \emph {et~al.}(2012)\citenamefont {Koller},
  \citenamefont {Tran},\ and\ \citenamefont {Blaha}}]{Koller2012:PRB}%
  \BibitemOpen
  \bibfield  {author} {\bibinfo {author} {\bibfnamefont {D.}~\bibnamefont
  {Koller}}, \bibinfo {author} {\bibfnamefont {F.}~\bibnamefont {Tran}}, \ and\
  \bibinfo {author} {\bibfnamefont {P.}~\bibnamefont {Blaha}},\ }\href
  {\doibase 10.1103/PhysRevB.85.155109} {\bibfield  {journal} {\bibinfo
  {journal} {Phys. Rev. B}\ }\textbf {\bibinfo {volume} {85}},\ \bibinfo
  {pages} {155109} (\bibinfo {year} {2012})}\BibitemShut {NoStop}%
\bibitem [{\citenamefont {Kim}\ \emph {et~al.}(2009)\citenamefont {Kim},
  \citenamefont {Hummer},\ and\ \citenamefont {Kresse}}]{Kim2009:PRB}%
  \BibitemOpen
  \bibfield  {author} {\bibinfo {author} {\bibfnamefont {Y.-S.}\ \bibnamefont
  {Kim}}, \bibinfo {author} {\bibfnamefont {K.}~\bibnamefont {Hummer}}, \ and\
  \bibinfo {author} {\bibfnamefont {G.}~\bibnamefont {Kresse}},\ }\href
  {\doibase 10.1103/PhysRevB.80.035203} {\bibfield  {journal} {\bibinfo
  {journal} {Phys. Rev. B}\ }\textbf {\bibinfo {volume} {80}},\ \bibinfo
  {pages} {035203} (\bibinfo {year} {2009})}\BibitemShut {NoStop}%
\bibitem [{\citenamefont {Luo}\ \emph {et~al.}(2009)\citenamefont {Luo},
  \citenamefont {Bester},\ and\ \citenamefont {Zunger}}]{Luo2009:PRL}%
  \BibitemOpen
  \bibfield  {author} {\bibinfo {author} {\bibfnamefont {J.-W.}\ \bibnamefont
  {Luo}}, \bibinfo {author} {\bibfnamefont {G.}~\bibnamefont {Bester}}, \ and\
  \bibinfo {author} {\bibfnamefont {A.}~\bibnamefont {Zunger}},\ }\href
  {\doibase 10.1103/PhysRevLett.102.056405} {\bibfield  {journal} {\bibinfo
  {journal} {Phys. Rev. Lett.}\ }\textbf {\bibinfo {volume} {102}},\ \bibinfo
  {pages} {056405} (\bibinfo {year} {2009})}\BibitemShut {NoStop}%
\bibitem [{\citenamefont {Perdew}\ \emph {et~al.}(1996)\citenamefont {Perdew},
  \citenamefont {Burke},\ and\ \citenamefont {Ernzerhof}}]{Perdew1996:PRL}%
  \BibitemOpen
  \bibfield  {author} {\bibinfo {author} {\bibfnamefont {J.~P.}\ \bibnamefont
  {Perdew}}, \bibinfo {author} {\bibfnamefont {K.}~\bibnamefont {Burke}}, \
  and\ \bibinfo {author} {\bibfnamefont {M.}~\bibnamefont {Ernzerhof}},\ }\href
  {\doibase 10.1103/PhysRevLett.77.3865} {\bibfield  {journal} {\bibinfo
  {journal} {Phys. Rev. Lett.}\ }\textbf {\bibinfo {volume} {77}},\ \bibinfo
  {pages} {3865} (\bibinfo {year} {1996})}\BibitemShut {NoStop}%
\bibitem [{\citenamefont {Koster}\ \emph {et~al.}(1963)\citenamefont {Koster},
  \citenamefont {Dimmock}, \citenamefont {Wheeler},\ and\ \citenamefont
  {Statz}}]{Koster1963:book}%
  \BibitemOpen
  \bibfield  {author} {\bibinfo {author} {\bibfnamefont {G.~F.}\ \bibnamefont
  {Koster}}, \bibinfo {author} {\bibfnamefont {J.~O.}\ \bibnamefont {Dimmock}},
  \bibinfo {author} {\bibfnamefont {R.~G.}\ \bibnamefont {Wheeler}}, \ and\
  \bibinfo {author} {\bibfnamefont {H.}~\bibnamefont {Statz}},\ }\href@noop {}
  {\emph {\bibinfo {title} {Properties of the thirty-two point groups}}}\
  (\bibinfo  {publisher} {Cambridge, MIT Press},\ \bibinfo {year}
  {1963})\BibitemShut {NoStop}%
\bibitem [{\citenamefont {Altmann}\ and\ \citenamefont
  {Herzig}(1994)}]{Altmann1994:book}%
  \BibitemOpen
  \bibfield  {author} {\bibinfo {author} {\bibfnamefont {S.}~\bibnamefont
  {Altmann}}\ and\ \bibinfo {author} {\bibfnamefont {P.}~\bibnamefont
  {Herzig}},\ }\href@noop {} {\emph {\bibinfo {title} {Point-Group Theory
  Tables}}}\ (\bibinfo  {publisher} {Oxford University, Oxford},\ \bibinfo
  {year} {1994})\BibitemShut {NoStop}%
\bibitem [{\citenamefont {Blaha}\ \emph {et~al.}(2013)\citenamefont {Blaha},
  \citenamefont {Schwarz}, \citenamefont {Madsen}, \citenamefont {Kvasnicka},\
  and\ \citenamefont {Luitz}}]{wien2k}%
  \BibitemOpen
  \bibfield  {author} {\bibinfo {author} {\bibfnamefont {P.}~\bibnamefont
  {Blaha}}, \bibinfo {author} {\bibfnamefont {K.}~\bibnamefont {Schwarz}},
  \bibinfo {author} {\bibfnamefont {G.~K.~H.}\ \bibnamefont {Madsen}}, \bibinfo
  {author} {\bibfnamefont {D.}~\bibnamefont {Kvasnicka}}, \ and\ \bibinfo
  {author} {\bibfnamefont {J.}~\bibnamefont {Luitz}},\ }\href
  {http://www.wien2k.at/} {\emph {\bibinfo {title} {Wien2k, An Augmented Plane
  Wave + Local Orbitals Program for Calculating Crystal Properties}}}\
  (\bibinfo  {publisher} {Vienna University of Technology},\ \bibinfo {year}
  {2013})\BibitemShut {NoStop}%
\bibitem [{\citenamefont {Kune\ifmmode~\check{s}\else \v{s}\fi{}}\ \emph
  {et~al.}(2001)\citenamefont {Kune\ifmmode~\check{s}\else \v{s}\fi{}},
  \citenamefont {Nov\'ak}, \citenamefont {Schmid}, \citenamefont {Blaha},\ and\
  \citenamefont {Schwarz}}]{Kunes2001:PRB}%
  \BibitemOpen
  \bibfield  {author} {\bibinfo {author} {\bibfnamefont {J.}~\bibnamefont
  {Kune\ifmmode~\check{s}\else \v{s}\fi{}}}, \bibinfo {author} {\bibfnamefont
  {P.}~\bibnamefont {Nov\'ak}}, \bibinfo {author} {\bibfnamefont
  {R.}~\bibnamefont {Schmid}}, \bibinfo {author} {\bibfnamefont
  {P.}~\bibnamefont {Blaha}}, \ and\ \bibinfo {author} {\bibfnamefont
  {K.}~\bibnamefont {Schwarz}},\ }\href {\doibase 10.1103/PhysRevB.64.153102}
  {\bibfield  {journal} {\bibinfo  {journal} {Phys. Rev. B}\ }\textbf {\bibinfo
  {volume} {64}},\ \bibinfo {pages} {153102} (\bibinfo {year}
  {2001})}\BibitemShut {NoStop}%
\bibitem [{\citenamefont {Singh}\ and\ \citenamefont
  {Nordstrom}(2006)}]{Singh2006:book}%
  \BibitemOpen
  \bibfield  {author} {\bibinfo {author} {\bibfnamefont {D.~J.}\ \bibnamefont
  {Singh}}\ and\ \bibinfo {author} {\bibfnamefont {L.}~\bibnamefont
  {Nordstrom}},\ }\href@noop {} {\emph {\bibinfo {title} {Planewaves,
  Pseudopotentials, and the LAPW Method}}}\ (\bibinfo  {publisher} {Springer
  US},\ \bibinfo {year} {2006})\BibitemShut {NoStop}%
\bibitem [{\citenamefont {De}\ and\ \citenamefont {Pryor}(2010)}]{De2010:PRB}%
  \BibitemOpen
  \bibfield  {author} {\bibinfo {author} {\bibfnamefont {A.}~\bibnamefont
  {De}}\ and\ \bibinfo {author} {\bibfnamefont {C.~E.}\ \bibnamefont {Pryor}},\
  }\href {\doibase 10.1103/PhysRevB.81.155210} {\bibfield  {journal} {\bibinfo
  {journal} {Phys. Rev. B}\ }\textbf {\bibinfo {volume} {81}},\ \bibinfo
  {pages} {155210} (\bibinfo {year} {2010})}\BibitemShut {NoStop}%
\bibitem [{\citenamefont {Dacal}\ and\ \citenamefont
  {Cantarero}(2014)}]{Dacal2014:MRE}%
  \BibitemOpen
  \bibfield  {author} {\bibinfo {author} {\bibfnamefont {L.~C.~O.}\
  \bibnamefont {Dacal}}\ and\ \bibinfo {author} {\bibfnamefont
  {A.}~\bibnamefont {Cantarero}},\ }\href {\doibase
  10.1088/2053-1591/1/1/015702} {\bibfield  {journal} {\bibinfo  {journal}
  {Mater. Res. Express}\ }\textbf {\bibinfo {volume} {1}},\ \bibinfo {pages}
  {015702} (\bibinfo {year} {2014})}\BibitemShut {NoStop}%
\bibitem [{\citenamefont {Furthmeier}\ \emph {et~al.}(2014)\citenamefont
  {Furthmeier}, \citenamefont {Dirnberger}, \citenamefont {Hubmann},
  \citenamefont {Bauer}, \citenamefont {Korn}, \citenamefont {Sch\"uller},
  \citenamefont {Zweck}, \citenamefont {Reiger},\ and\ \citenamefont
  {Bougeard}}]{Furthmeier2014:APL}%
  \BibitemOpen
  \bibfield  {author} {\bibinfo {author} {\bibfnamefont {S.}~\bibnamefont
  {Furthmeier}}, \bibinfo {author} {\bibfnamefont {F.}~\bibnamefont
  {Dirnberger}}, \bibinfo {author} {\bibfnamefont {J.}~\bibnamefont {Hubmann}},
  \bibinfo {author} {\bibfnamefont {B.}~\bibnamefont {Bauer}}, \bibinfo
  {author} {\bibfnamefont {T.}~\bibnamefont {Korn}}, \bibinfo {author}
  {\bibfnamefont {C.}~\bibnamefont {Sch\"uller}}, \bibinfo {author}
  {\bibfnamefont {J.}~\bibnamefont {Zweck}}, \bibinfo {author} {\bibfnamefont
  {E.}~\bibnamefont {Reiger}}, \ and\ \bibinfo {author} {\bibfnamefont
  {D.}~\bibnamefont {Bougeard}},\ }\href {\doibase 10.1063/1.4903482}
  {\bibfield  {journal} {\bibinfo  {journal} {Applied Physics Letters}\
  }\textbf {\bibinfo {volume} {105}},\ \bibinfo {pages} {222109} (\bibinfo
  {year} {2014})}\BibitemShut {NoStop}%
\bibitem [{\citenamefont {Ketterer}\ \emph {et~al.}(2011)\citenamefont
  {Ketterer}, \citenamefont {Heiss}, \citenamefont {Livrozet}, \citenamefont
  {Rudolph}, \citenamefont {Reiger},\ and\ \citenamefont {Fontcuberta~i
  Morral}}]{Ketterer2011:PRB}%
  \BibitemOpen
  \bibfield  {author} {\bibinfo {author} {\bibfnamefont {B.}~\bibnamefont
  {Ketterer}}, \bibinfo {author} {\bibfnamefont {M.}~\bibnamefont {Heiss}},
  \bibinfo {author} {\bibfnamefont {M.~J.}\ \bibnamefont {Livrozet}}, \bibinfo
  {author} {\bibfnamefont {A.}~\bibnamefont {Rudolph}}, \bibinfo {author}
  {\bibfnamefont {E.}~\bibnamefont {Reiger}}, \ and\ \bibinfo {author}
  {\bibfnamefont {A.}~\bibnamefont {Fontcuberta~i Morral}},\ }\href {\doibase
  10.1103/PhysRevB.83.125307} {\bibfield  {journal} {\bibinfo  {journal} {Phys.
  Rev. B}\ }\textbf {\bibinfo {volume} {83}},\ \bibinfo {pages} {125307}
  (\bibinfo {year} {2011})}\BibitemShut {NoStop}%
\bibitem [{\citenamefont {M{\"o}ller}\ \emph {et~al.}(2012)\citenamefont
  {M{\"o}ller}, \citenamefont {Jr}, \citenamefont {Cantarero}, \citenamefont
  {Chiaramonte}, \citenamefont {Cotta},\ and\ \citenamefont
  {Iikawa}}]{Moller2012:NT}%
  \BibitemOpen
  \bibfield  {author} {\bibinfo {author} {\bibfnamefont {M.}~\bibnamefont
  {M{\"o}ller}}, \bibinfo {author} {\bibfnamefont {M.~M. d.~L.}\ \bibnamefont
  {Jr}}, \bibinfo {author} {\bibfnamefont {A.}~\bibnamefont {Cantarero}},
  \bibinfo {author} {\bibfnamefont {T.}~\bibnamefont {Chiaramonte}}, \bibinfo
  {author} {\bibfnamefont {M.~A.}\ \bibnamefont {Cotta}}, \ and\ \bibinfo
  {author} {\bibfnamefont {F.}~\bibnamefont {Iikawa}},\ }\href {\doibase
  10.1088/0957-4484/23/37/375704} {\bibfield  {journal} {\bibinfo  {journal}
  {Nanotechnology}\ }\textbf {\bibinfo {volume} {23}},\ \bibinfo {pages}
  {375704} (\bibinfo {year} {2012})}\BibitemShut {NoStop}%
\bibitem [{\citenamefont {Joulli\'e}\ \emph {et~al.}(1981)\citenamefont
  {Joulli\'e}, \citenamefont {Eddin},\ and\ \citenamefont
  {Girault}}]{Joullie1981:PRB}%
  \BibitemOpen
  \bibfield  {author} {\bibinfo {author} {\bibfnamefont {A.}~\bibnamefont
  {Joulli\'e}}, \bibinfo {author} {\bibfnamefont {A.~Z.}\ \bibnamefont
  {Eddin}}, \ and\ \bibinfo {author} {\bibfnamefont {B.}~\bibnamefont
  {Girault}},\ }\href {\doibase 10.1103/PhysRevB.23.928} {\bibfield  {journal}
  {\bibinfo  {journal} {Phys. Rev. B}\ }\textbf {\bibinfo {volume} {23}},\
  \bibinfo {pages} {928} (\bibinfo {year} {1981})}\BibitemShut {NoStop}%
\bibitem [{\citenamefont {Yeh}\ \emph {et~al.}(1994)\citenamefont {Yeh},
  \citenamefont {Wei},\ and\ \citenamefont {Zunger}}]{Yeh1994:PRB}%
  \BibitemOpen
  \bibfield  {author} {\bibinfo {author} {\bibfnamefont {C.-Y.}\ \bibnamefont
  {Yeh}}, \bibinfo {author} {\bibfnamefont {S.-H.}\ \bibnamefont {Wei}}, \ and\
  \bibinfo {author} {\bibfnamefont {A.}~\bibnamefont {Zunger}},\ }\href
  {\doibase 10.1103/PhysRevB.50.2715} {\bibfield  {journal} {\bibinfo
  {journal} {Phys. Rev. B}\ }\textbf {\bibinfo {volume} {50}},\ \bibinfo
  {pages} {2715} (\bibinfo {year} {1994})}\BibitemShut {NoStop}%
\bibitem [{\citenamefont {Birman}(1959)}]{Birman1959:PRL}%
  \BibitemOpen
  \bibfield  {author} {\bibinfo {author} {\bibfnamefont {J.~L.}\ \bibnamefont
  {Birman}},\ }\href {\doibase 10.1103/PhysRevLett.2.157} {\bibfield  {journal}
  {\bibinfo  {journal} {Phys. Rev. Lett.}\ }\textbf {\bibinfo {volume} {2}},\
  \bibinfo {pages} {157} (\bibinfo {year} {1959})}\BibitemShut {NoStop}%
\bibitem [{\citenamefont {Peng}\ \emph {et~al.}(2012)\citenamefont {Peng},
  \citenamefont {Jabeen}, \citenamefont {Jusserand}, \citenamefont {Harmand},
  \citenamefont {Bernard}, \citenamefont {Harmand}, \citenamefont {Jusserand},
  \citenamefont {Jabeen},\ and\ \citenamefont {Peng}}]{Peng2012:APL}%
  \BibitemOpen
  \bibfield  {author} {\bibinfo {author} {\bibfnamefont {W.}~\bibnamefont
  {Peng}}, \bibinfo {author} {\bibfnamefont {F.}~\bibnamefont {Jabeen}},
  \bibinfo {author} {\bibfnamefont {B.}~\bibnamefont {Jusserand}}, \bibinfo
  {author} {\bibfnamefont {J.~C.}\ \bibnamefont {Harmand}}, \bibinfo {author}
  {\bibfnamefont {M.}~\bibnamefont {Bernard}}, \bibinfo {author} {\bibfnamefont
  {J.~C.}\ \bibnamefont {Harmand}}, \bibinfo {author} {\bibfnamefont
  {B.}~\bibnamefont {Jusserand}}, \bibinfo {author} {\bibfnamefont
  {F.}~\bibnamefont {Jabeen}}, \ and\ \bibinfo {author} {\bibfnamefont
  {W.}~\bibnamefont {Peng}},\ }\href {\doibase 10.1063/1.3684837} {\bibfield
  {journal} {\bibinfo  {journal} {Applied Physics Letters}\ }\textbf {\bibinfo
  {volume} {100}},\ \bibinfo {pages} {073102} (\bibinfo {year}
  {2012})}\BibitemShut {NoStop}%
\bibitem [{\citenamefont {Kusch}\ \emph {et~al.}(2012)\citenamefont {Kusch},
  \citenamefont {Breuer}, \citenamefont {Ramsteiner}, \citenamefont {Geelhaar},
  \citenamefont {Riechert},\ and\ \citenamefont {Reich}}]{Kusch2012:PRB}%
  \BibitemOpen
  \bibfield  {author} {\bibinfo {author} {\bibfnamefont {P.}~\bibnamefont
  {Kusch}}, \bibinfo {author} {\bibfnamefont {S.}~\bibnamefont {Breuer}},
  \bibinfo {author} {\bibfnamefont {M.}~\bibnamefont {Ramsteiner}}, \bibinfo
  {author} {\bibfnamefont {L.}~\bibnamefont {Geelhaar}}, \bibinfo {author}
  {\bibfnamefont {H.}~\bibnamefont {Riechert}}, \ and\ \bibinfo {author}
  {\bibfnamefont {S.}~\bibnamefont {Reich}},\ }\href {\doibase
  10.1103/PhysRevB.86.075317} {\bibfield  {journal} {\bibinfo  {journal} {Phys.
  Rev. B}\ }\textbf {\bibinfo {volume} {86}},\ \bibinfo {pages} {075317}
  (\bibinfo {year} {2012})}\BibitemShut {NoStop}%
\bibitem [{\citenamefont {Peng}\ and\ \citenamefont
  {Copple}(2013)}]{Peng2013:PRB}%
  \BibitemOpen
  \bibfield  {author} {\bibinfo {author} {\bibfnamefont {X.}~\bibnamefont
  {Peng}}\ and\ \bibinfo {author} {\bibfnamefont {A.}~\bibnamefont {Copple}},\
  }\href {\doibase 10.1103/PhysRevB.87.115308} {\bibfield  {journal} {\bibinfo
  {journal} {Phys. Rev. B}\ }\textbf {\bibinfo {volume} {87}},\ \bibinfo
  {pages} {115308} (\bibinfo {year} {2013})}\BibitemShut {NoStop}%
\bibitem [{\citenamefont {Zanolli}\ \emph
  {et~al.}(2007{\natexlab{b}})\citenamefont {Zanolli}, \citenamefont {Pistol},
  \citenamefont {Fr{\"o}berg},\ and\ \citenamefont
  {Samuelson}}]{Zanolli2007:JOP}%
  \BibitemOpen
  \bibfield  {author} {\bibinfo {author} {\bibfnamefont {Z.}~\bibnamefont
  {Zanolli}}, \bibinfo {author} {\bibfnamefont {M.-E.}\ \bibnamefont {Pistol}},
  \bibinfo {author} {\bibfnamefont {L.~E.}\ \bibnamefont {Fr{\"o}berg}}, \ and\
  \bibinfo {author} {\bibfnamefont {L.}~\bibnamefont {Samuelson}},\ }\href
  {http://stacks.iop.org/0953-8984/19/i=29/a=295219} {\bibfield  {journal}
  {\bibinfo  {journal} {Journal of Physics: Condensed Matter}\ }\textbf
  {\bibinfo {volume} {19}},\ \bibinfo {pages} {295219} (\bibinfo {year}
  {2007}{\natexlab{b}})}\BibitemShut {NoStop}%
\bibitem [{\citenamefont {Casella}(1959)}]{Casella1959:PR}%
  \BibitemOpen
  \bibfield  {author} {\bibinfo {author} {\bibfnamefont {R.~C.}\ \bibnamefont
  {Casella}},\ }\href {\doibase 10.1103/PhysRev.114.1514} {\bibfield  {journal}
  {\bibinfo  {journal} {Phys. Rev.}\ }\textbf {\bibinfo {volume} {114}},\
  \bibinfo {pages} {1514} (\bibinfo {year} {1959})}\BibitemShut {NoStop}%
\bibitem [{\citenamefont {Elliott}(1954)}]{Elliot1954:PR}%
  \BibitemOpen
  \bibfield  {author} {\bibinfo {author} {\bibfnamefont {R.~J.}\ \bibnamefont
  {Elliott}},\ }\href {\doibase 10.1103/PhysRev.96.280} {\bibfield  {journal}
  {\bibinfo  {journal} {Phys. Rev.}\ }\textbf {\bibinfo {volume} {96}},\
  \bibinfo {pages} {280} (\bibinfo {year} {1954})}\BibitemShut {NoStop}%
\bibitem [{\citenamefont {Cardona}\ \emph {et~al.}(1986)\citenamefont
  {Cardona}, \citenamefont {Christensen},\ and\ \citenamefont
  {Fasol}}]{Cardona1986:PRL}%
  \BibitemOpen
  \bibfield  {author} {\bibinfo {author} {\bibfnamefont {M.}~\bibnamefont
  {Cardona}}, \bibinfo {author} {\bibfnamefont {N.~E.}\ \bibnamefont
  {Christensen}}, \ and\ \bibinfo {author} {\bibfnamefont {G.}~\bibnamefont
  {Fasol}},\ }\href {\doibase 10.1103/PhysRevLett.56.2831} {\bibfield
  {journal} {\bibinfo  {journal} {Phys. Rev. Lett.}\ }\textbf {\bibinfo
  {volume} {56}},\ \bibinfo {pages} {2831} (\bibinfo {year}
  {1986})}\BibitemShut {NoStop}%
\bibitem [{\citenamefont {Krich}\ and\ \citenamefont
  {Halperin}(2007)}]{Krich2007:PRL}%
  \BibitemOpen
  \bibfield  {author} {\bibinfo {author} {\bibfnamefont {J.~J.}\ \bibnamefont
  {Krich}}\ and\ \bibinfo {author} {\bibfnamefont {B.~I.}\ \bibnamefont
  {Halperin}},\ }\href {\doibase 10.1103/PhysRevLett.98.226802} {\bibfield
  {journal} {\bibinfo  {journal} {Phys. Rev. Lett.}\ }\textbf {\bibinfo
  {volume} {98}},\ \bibinfo {pages} {226802} (\bibinfo {year}
  {2007})}\BibitemShut {NoStop}%
\bibitem [{\citenamefont {Rashba}(1960)}]{Rashba1960:SPSS}%
  \BibitemOpen
  \bibfield  {author} {\bibinfo {author} {\bibfnamefont {E.~I.}\ \bibnamefont
  {Rashba}},\ }\href@noop {} {\bibfield  {journal} {\bibinfo  {journal} {Sov.
  Phys. Sol. State}\ }\textbf {\bibinfo {volume} {2}},\ \bibinfo {pages} {1109}
  (\bibinfo {year} {1960})}\BibitemShut {NoStop}%
\bibitem [{\citenamefont {Lew Yan~Voon}\ \emph {et~al.}(1996)\citenamefont {Lew
  Yan~Voon}, \citenamefont {Willatzen}, \citenamefont {Cardona},\ and\
  \citenamefont {Christensen}}]{LewYanVoon1996:PRB}%
  \BibitemOpen
  \bibfield  {author} {\bibinfo {author} {\bibfnamefont {L.~C.}\ \bibnamefont
  {Lew Yan~Voon}}, \bibinfo {author} {\bibfnamefont {M.}~\bibnamefont
  {Willatzen}}, \bibinfo {author} {\bibfnamefont {M.}~\bibnamefont {Cardona}},
  \ and\ \bibinfo {author} {\bibfnamefont {N.~E.}\ \bibnamefont
  {Christensen}},\ }\href {\doibase 10.1103/PhysRevB.53.10703} {\bibfield
  {journal} {\bibinfo  {journal} {Phys. Rev. B}\ }\textbf {\bibinfo {volume}
  {53}},\ \bibinfo {pages} {10703} (\bibinfo {year} {1996})}\BibitemShut
  {NoStop}%
\bibitem [{\citenamefont {Hopfield}(1961)}]{Hopfield1961:JAP}%
  \BibitemOpen
  \bibfield  {author} {\bibinfo {author} {\bibfnamefont {J.~J.}\ \bibnamefont
  {Hopfield}},\ }\href {\doibase 10.1063/1.1777059} {\bibfield  {journal}
  {\bibinfo  {journal} {Journal of Applied Physics}\ }\textbf {\bibinfo
  {volume} {32}},\ \bibinfo {pages} {2277} (\bibinfo {year}
  {1961})}\BibitemShut {NoStop}%
\bibitem [{\citenamefont {Mahan}\ and\ \citenamefont
  {Hopfield}(1964)}]{Mahan1964:PR}%
  \BibitemOpen
  \bibfield  {author} {\bibinfo {author} {\bibfnamefont {G.~D.}\ \bibnamefont
  {Mahan}}\ and\ \bibinfo {author} {\bibfnamefont {J.~J.}\ \bibnamefont
  {Hopfield}},\ }\href {\doibase 10.1103/PhysRev.135.A428} {\bibfield
  {journal} {\bibinfo  {journal} {Phys. Rev.}\ }\textbf {\bibinfo {volume}
  {135}},\ \bibinfo {pages} {A428} (\bibinfo {year} {1964})}\BibitemShut
  {NoStop}%
\bibitem [{\citenamefont {H{\"u}mmer}\ \emph {et~al.}(1978)\citenamefont
  {H{\"u}mmer}, \citenamefont {Helbig},\ and\ \citenamefont
  {Baumg{\"a}rtner}}]{Hummer1978:PSSB}%
  \BibitemOpen
  \bibfield  {author} {\bibinfo {author} {\bibfnamefont {K.}~\bibnamefont
  {H{\"u}mmer}}, \bibinfo {author} {\bibfnamefont {R.}~\bibnamefont {Helbig}},
  \ and\ \bibinfo {author} {\bibfnamefont {M.}~\bibnamefont
  {Baumg{\"a}rtner}},\ }\href {\doibase 10.1002/pssb.2220860211} {\bibfield
  {journal} {\bibinfo  {journal} {phys. stat. sol. (b)}\ }\textbf {\bibinfo
  {volume} {86}},\ \bibinfo {pages} {527} (\bibinfo {year} {1978})}\BibitemShut
  {NoStop}%
\bibitem [{\citenamefont {Broser}\ and\ \citenamefont
  {Rosenzweig}(1979)}]{Broser1979:PSSB}%
  \BibitemOpen
  \bibfield  {author} {\bibinfo {author} {\bibfnamefont {I.}~\bibnamefont
  {Broser}}\ and\ \bibinfo {author} {\bibfnamefont {M.}~\bibnamefont
  {Rosenzweig}},\ }\href {\doibase 10.1002/pssb.2220950116} {\bibfield
  {journal} {\bibinfo  {journal} {phys. stat. sol. (b)}\ }\textbf {\bibinfo
  {volume} {95}},\ \bibinfo {pages} {141} (\bibinfo {year} {1979})}\BibitemShut
  {NoStop}%
\bibitem [{\citenamefont {Koteles}\ and\ \citenamefont
  {Winterling}(1980)}]{Koteles1980:PRL}%
  \BibitemOpen
  \bibfield  {author} {\bibinfo {author} {\bibfnamefont {E.~S.}\ \bibnamefont
  {Koteles}}\ and\ \bibinfo {author} {\bibfnamefont {G.}~\bibnamefont
  {Winterling}},\ }\href {\doibase 10.1103/PhysRevLett.44.948} {\bibfield
  {journal} {\bibinfo  {journal} {Phys. Rev. Lett.}\ }\textbf {\bibinfo
  {volume} {44}},\ \bibinfo {pages} {948} (\bibinfo {year} {1980})}\BibitemShut
  {NoStop}%
\bibitem [{\citenamefont {Fu}\ and\ \citenamefont {Wu}(2008)}]{Fu2008:JAP}%
  \BibitemOpen
  \bibfield  {author} {\bibinfo {author} {\bibfnamefont {J.~Y.}\ \bibnamefont
  {Fu}}\ and\ \bibinfo {author} {\bibfnamefont {M.~W.}\ \bibnamefont {Wu}},\
  }\href {\doibase 10.1063/1.3018600} {\bibfield  {journal} {\bibinfo
  {journal} {Journal of Applied Physics}\ }\textbf {\bibinfo {volume} {104}},\
  \bibinfo {pages} {093712} (\bibinfo {year} {2008})}\BibitemShut {NoStop}%
\bibitem [{\citenamefont {Wang}\ \emph {et~al.}(2007)\citenamefont {Wang},
  \citenamefont {Wu}, \citenamefont {Tsay}, \citenamefont {Gau}, \citenamefont
  {Lo}, \citenamefont {Kao}, \citenamefont {Jang},\ and\ \citenamefont
  {Chiang}}]{Wang2007:APL}%
  \BibitemOpen
  \bibfield  {author} {\bibinfo {author} {\bibfnamefont {W.-T.}\ \bibnamefont
  {Wang}}, \bibinfo {author} {\bibfnamefont {C.~L.}\ \bibnamefont {Wu}},
  \bibinfo {author} {\bibfnamefont {S.~F.}\ \bibnamefont {Tsay}}, \bibinfo
  {author} {\bibfnamefont {M.~H.}\ \bibnamefont {Gau}}, \bibinfo {author}
  {\bibfnamefont {I.}~\bibnamefont {Lo}}, \bibinfo {author} {\bibfnamefont
  {H.~F.}\ \bibnamefont {Kao}}, \bibinfo {author} {\bibfnamefont {D.~J.}\
  \bibnamefont {Jang}}, \ and\ \bibinfo {author} {\bibfnamefont {J.-C.}\
  \bibnamefont {Chiang}},\ }\href {\doibase 10.1063/1.2775038} {\bibfield
  {journal} {\bibinfo  {journal} {Applied Physics Letters}\ }\textbf {\bibinfo
  {volume} {91}},\ \bibinfo {pages} {082110} (\bibinfo {year}
  {2007})}\BibitemShut {NoStop}%
\bibitem [{\citenamefont {Vurgaftman}\ \emph {et~al.}(2001)\citenamefont
  {Vurgaftman}, \citenamefont {Meyer},\ and\ \citenamefont
  {Ram-Mohan}}]{Vurgaftman2001:JAP}%
  \BibitemOpen
  \bibfield  {author} {\bibinfo {author} {\bibfnamefont {I.}~\bibnamefont
  {Vurgaftman}}, \bibinfo {author} {\bibfnamefont {J.~R.}\ \bibnamefont
  {Meyer}}, \ and\ \bibinfo {author} {\bibfnamefont {L.~R.}\ \bibnamefont
  {Ram-Mohan}},\ }\href {\doibase 10.1063/1.1368156} {\bibfield  {journal}
  {\bibinfo  {journal} {Journal of Applied Physics}\ }\textbf {\bibinfo
  {volume} {89}},\ \bibinfo {pages} {5815} (\bibinfo {year}
  {2001})}\BibitemShut {NoStop}%
\bibitem [{\citenamefont {Madelung}(2004)}]{Madelung2004:book}%
  \BibitemOpen
  \bibfield  {author} {\bibinfo {author} {\bibfnamefont {P.~D.~O.}\
  \bibnamefont {Madelung}},\ }\href
  {http://link.springer.com/chapter/10.1007/978-3-642-18865-7_1} {\emph
  {\bibinfo {title} {Semiconductors: Data Handbook}}}\ (\bibinfo  {publisher}
  {Springer Berlin Heidelberg},\ \bibinfo {year} {2004})\BibitemShut {NoStop}%
\bibitem [{\citenamefont {Bublik}\ \emph {et~al.}(1982)\citenamefont {Bublik},
  \citenamefont {Wilke},\ and\ \citenamefont {Pereversev}}]{Bublik1982:pssa}%
  \BibitemOpen
  \bibfield  {author} {\bibinfo {author} {\bibfnamefont {V.~T.}\ \bibnamefont
  {Bublik}}, \bibinfo {author} {\bibfnamefont {J.}~\bibnamefont {Wilke}}, \
  and\ \bibinfo {author} {\bibfnamefont {A.~T.}\ \bibnamefont {Pereversev}},\
  }\href {\doibase 10.1002/pssa.2210730270} {\bibfield  {journal} {\bibinfo
  {journal} {phys. stat. sol. (a)}\ }\textbf {\bibinfo {volume} {73}},\
  \bibinfo {pages} {K271} (\bibinfo {year} {1982})}\BibitemShut {NoStop}%
\bibitem [{\citenamefont {Nilsen}\ \emph {et~al.}(2010)\citenamefont {Nilsen},
  \citenamefont {Breivik}, \citenamefont {Myrvågnes}, \citenamefont {Nilsen},
  \citenamefont {Breivik}, \citenamefont {Fimland},\ and\ \citenamefont
  {Fimland}}]{Nilsen2010:JVST}%
  \BibitemOpen
  \bibfield  {author} {\bibinfo {author} {\bibfnamefont {T.~A.}\ \bibnamefont
  {Nilsen}}, \bibinfo {author} {\bibfnamefont {M.}~\bibnamefont {Breivik}},
  \bibinfo {author} {\bibfnamefont {G.}~\bibnamefont {Myrvågnes}}, \bibinfo
  {author} {\bibfnamefont {T.~A.}\ \bibnamefont {Nilsen}}, \bibinfo {author}
  {\bibfnamefont {M.}~\bibnamefont {Breivik}}, \bibinfo {author} {\bibfnamefont
  {B.-O.}\ \bibnamefont {Fimland}}, \ and\ \bibinfo {author} {\bibfnamefont
  {B.-O.}\ \bibnamefont {Fimland}},\ }\href {\doibase 10.1116/1.3336341}
  {\bibfield  {journal} {\bibinfo  {journal} {Journal of Vacuum Science \&
  Technology B}\ }\textbf {\bibinfo {volume} {28}},\ \bibinfo {pages} {C3I17}
  (\bibinfo {year} {2010})}\BibitemShut {NoStop}%
\end{thebibliography}%
\end{document}